\documentclass[aps,apl,twocolumn,floatfix,superscriptaddress]{revtex4}
\usepackage{graphicx}
\usepackage{amsmath}
\usepackage{hyperref}
\usepackage{dcolumn}
\usepackage{bm}
\usepackage[dvips]{color}
\graphicspath{{JAP/}}

\newcommand{\degr}{$^{\circ}$}
\newcommand{\etal} {\textit{et al.}}
\newcommand{\ie} {\textit{i.e.}}
\newcommand {\eg} {\textit{e.g.}}

\newcommand{\fzd} {Institute of Ion Beam Physics and Materials Research, Forschungszentrum Dresden-Rossendorf, P.O. Box 510119, 01314 Dresden, Germany}

\newcommand{\chim} {$\chi_{min}$}

\begin{document}

\preprint{APS/123-QED}

\title{Fe implanted ZnO: magnetic precipitates versus dilution}

\author{Shengqiang~Zhou}
\email[Electronic address: ]{S.Zhou@fzd.de}
\author{K.~Potzger}
\author{G.~Talut}
\author{H~Reuther}
\affiliation{\fzd}
\author{N. Volbers}
\affiliation{Justus-Liebig-University Gie$\ss$en, I. Physic
Insititute, Heinrich-Buff-Ring 16, 35392 Gie$\ss$en, Germany}
\author{M. Lorenz}
\affiliation{Universit\"{a}t Leipzig, Fakult\"{a}t f\"{u}r Physik
und Geowissenschaften, Institut f\"{u}r Experimentelle Physik II,
Linn\'{e}str. 5, D-04103 Leipzig, Germany }
\author{J.~von Borany}
\author{R.~Gr\"{o}tzschel}
\author{W.~Skorupa}
\author{M.~Helm}
\author{J.~Fassbender}
\affiliation{\fzd}
\author{T. Herrmannsd\"{o}rfer}
\affiliation{Hochfeld-Magnetlabor Dresden (HLD), Forschungszentrum
Dresden-Rossendorf, P.O. Box 510119, 01314 Dresden, Germany}
\begin{abstract}

Nowadays ferromagnetism is often found in potential diluted
magnetic semiconductor systems. However, many authors argue that
the observed ferromagnetism stems from ferromagnetic precipitates
or spinodal decomposition rather than from carrier mediated
magnetic impurities, as required for a diluted magnetic
semiconductor. In the present paper we answer this question for
Fe-implanted ZnO single crystals comprehensively. Different
implantation fluences and temperatures and post-implantation
annealing temperatures have been chosen in order to evaluate the
structural and magnetic properties over a wide range of
parameters. Three different regimes with respect to the Fe
concentration and the process temperature are found: 1) Disperse
Fe$^{2+}$ and Fe$^{3+}$ at low Fe concentrations and low
processing temperatures, 2) FeZn$_2$O$_4$ at very high processing
temperatures and 3) an intermediate regime with a co-existence of
metallic Fe (Fe$^0$) and ionic Fe (Fe$^{2+}$ and Fe$^{3+}$).
Ferromagnetism is only observed in the latter two cases, where
inverted ZnFe$_2$O$_4$ and $\alpha$-Fe nanocrystals are the origin
of the observed ferromagnetic behavior, respectively. The ionic Fe
in the last case could contribute to a carrier mediated coupling.
However, their separation is too large to couple ferromagnetically
due to the lack of p-type carrier. For comparison investigations
of Fe-implanted epitaxial ZnO thin films are presented.
\end{abstract}
\maketitle
\pagebreak

\section{Introduction}\label{section:introduction}

Recently, considerable interest has been paid to "spintronics",
where the spin degree of freedom of the electron carries
information in the device. One of the material systems to realize
this function are diluted magnetic semiconductors (DMS). In DMS
materials, transition or rare earth metal ions are substituted
onto cation sites and are coupled with the free carriers to yield
ferromagnetism via indirect interaction. In 2000, Dietl
\etal~\cite{dietl00} proposed the mean-field Zener model to
understand the ferromagnetism in DMS materials. It has been
successfully used to describe the magnetic coupling in (Ga,Mn)As
and (Zn,Mn)Te materials. This model predicts that wide bandgap
semiconductors (GaN and ZnO) doped with Mn exhibit ordering
temperatures above 300 K, provided that a sufficiently large hole
density can be achieved (10$^{20}$ cm$^{-3}$). Sato \etal~used the
Korringa-Kohn-Rostoker Green function method based on the local
density approximation of density functional theory to calculate
the properties of n-type ZnO doped with the 3$d$ TM ions (V, Cr,
Mn, Fe, Co, and Ni) \cite{sato_ZnO}. The ferromagnetic state, with
a T$_C$ of around 2000 K, is predicted to be favourable for V, Cr,
Fe, Co, and Ni in ZnO while Mn-doped ZnO is antiferromagnetic.
These predictions have largely boosted intensive experimental
activity on transition metal doped GaN and ZnO. A large number of
research groups reported the experimental observation of
ferromagnetism in TM (from Sc to Ni) doped ZnO
\cite{angadi06}\cite{heo04}\cite{hong05}\cite{hongv}\cite{ip03}\cite{jung02}\cite{pol04}\cite{tuan04}\cite{venk04}
fabricated by various methods including ion implantation. In
contrast to these results, other groups reported the observations
of antiferromagnetism \cite{boul05}\cite{yin06}\cite{sati:137204},
spin-glass behavior \cite{fukumura01}\cite{jin01}, and
paramagnetism \cite{yin06}\cite{rao05}\cite{zhang06} in TM-doped
ZnO. Recently it was also found that nanoscale precipitates can
substantially contribute to the ferromagnetic properties
\cite{norton03}\cite{park04}\cite{kund04}\cite{shim05}\cite{pot06fe}\cite{shin06}\cite{talut06}\cite{zhou06}\cite{zhou07JPD}.
One method to introduce magnetic dopants into ZnO is ion
implantation. It has several advantages, namely the
reproducibility, the precise control of the ion fluence, the use
of an isotopically pure beam, and the possibility to overcome the
solubility limit. The major drawback of ion implantation is the
generation of structural defects in the host lattice. However, in
several studies it has been demonstrated that ZnO exhibits a high
amorphization threshold. Therefore ion implantation is widely used
to dope ZnO with transition metal ions. Ref. \cite{hebard04} gives
a review on transition metal ion implantation into ZnO.

Hydrothermal growth is one of the major methods to fabricate high
quality ZnO single crystals \cite{ohshima04}\cite{izyumskaya07}.
Hydrothermal grown ZnO single crystals have been widely used for
photodiodes \cite{endo:121906}, light emitters \cite{gao:123125},
DMS \cite{pot06fe}\cite{li:112507}\cite{fenwick:64741Q}, and
substrates for homoepitaxial growth of ZnO films
\cite{matsui:2454}.

The present paper is dedicated to a comprehensive investigation of
the structural and magnetic properties of Fe implanted ZnO bulk
crystals grown by the hydrothermal method. Different implantation parameters, \ie~ion fluence, ion energy,
and implantation temperature, were varied. Three charge states or
occupied sites, \ie~metallic Fe and monodispersed Fe$^{2+}$ and
Fe$^{3+}$ and Fe$^{3+}$ in Zn-ferrites, are identified. Metallic
Fe nanocrystals (NCs) form after implantation at high fluence and
high temperature. They are the major contribution to the measured
ferromagnetism. The difference between ZnO single crystals and
epitaxial thin films upon the same implantation, and the
difference between the high (623 K) and the low (253 K)
temperature implanted samples subject to the same annealing, will
be discussed.

Actually the phase separation, namely MnAs precipitates, in
(Ga,Mn)As (the most well understood DMS material), has been
intensively investigated \cite{ohnohb}. Ferromagnetic MnAs
precipitates are epitaxially embedded inside the GaAs matrix, and
exhibit interesting magneto-transport properties
\cite{wellmann98}\cite{yuldashev01}\cite{ramsteiner02}\cite{yokoyama06}.
Sato and Katayama-Yoshida \cite{sato05}\cite{katayama07}
calculated the chemical pair interaction between two magnetic
impurities in DMS materials. A strong attractive interaction
between magnetic impurities has been found, which accelerates the
spinodal nano-decomposition under thermal non-equilibrium crystal
growth conditions. Moreover the attractive interaction in
(Ga,Mn)As is one order of magnitude weaker than that in (Ga,Mn)N
(wide bandgap semiconductor). Therefore, phase separation in TM
doped wide bandgap semiconductors is highly expected. In addition
to those investigations of MnAs/GaAs hybrids, some interesting
magneto-transport properties are demonstrated, \eg~anomalous hall
effect, and giant magnetoresistance, for magnetic NCs embedded
inside ZnS \cite{liu:092507} and Ge \cite{jamet06}, respectively.
Therefore, Fe NCs embedded inside ZnO, which are granular magnetic
nano-precipitates inside a semiconductor (granular magnetic
semiconductor, GMS), could have some potential applications in
future nano-spintronics \cite{katayama07}\cite{dietlmat}.

The paper is organized as follows. First, all the experimental
methods employed will be described. Then the results will be
separated according to the physical phenomena as follows: lattice
damage and recovering, the distribution of implanted Fe, the
formation of precipitates (metallic Fe, or Zn-ferrites), the
charge state of Fe, the ferromagnetic properties, and the Fe
implanted epitaxial ZnO films. In the discussion part, we sketch a
phase diagram of Fe in ZnO, and apply a model to explain the Fe
nanocrystal aggregation. Moreover the reason for the absence of
ferromagnetism in ionic Fe diluted ZnO is discussed.

\section{Experiments} \label{section:experiments}
Commercial ZnO bulk crystals were implanted with $^{57}$Fe ions at
temperatures ranging from 253 K to 623 K with fluences from
$0.1\times10^{16}$ cm$^{-2}$ to $8\times10^{16}$ cm$^{-2}$. The
implantation energy was 180 keV, which results in a projected
range of $R_P=89\pm29$ nm, and a maximum atomic concentration from
0.14\% to 11\% (TRIM code \cite{trim}). For comparison, epitaxial
ZnO thin films grown on Al$_2$O$_3$ by pulsed laser deposition
were implanted with $^{57}$Fe at selected implantation parameters
(623 K, $4\times10^{16}$ cm$^{-2}$). Three sample series are investigated and listed in Table
\ref{tab:FeZnO_sample}.

The lattice damage induced by implantation was evaluated by
Rutherford backscattering/channeling spectrometry (RBS/C). The
RBS/C spectra were collected with a collimated 1.7 MeV He$^+$ beam
at a backscattering angle of 170\degr. The sample was mounted on a
three-axis goniometer with a precision of 0.01\degr. The
channeling spectra were collected by aligning the sample to make
the impinging He$^+$ beam parallel with the ZnO$<$0001$> $ axis.
\chim~is the channeling minimum yield in RBS/C, which is the ratio
of the backscattering yield at channeling condition to that for a
random beam incidence \cite{chuwk}. Therefore, \chim~labels the
lattice disordering degree upon implantation. An amorphous sample
exhibits a \chim~of 100\%, while a perfect single crystal
corresponds to a \chim~of 1-2\%.

The Fe distributions were investigated by secondary ion mass
spectrometry (SIMS), using a Riber MIQ-256 system with oxygen
primary ions of 6 kV and monitoring positive secondary ions. The
depth scale was calibrated by measuring the sputtered crater via
profilometry. The absolute concentration was determined by
calculating the sensitivity factors from the low fluence implants.

Structural analysis was achieved both by synchrotron radiation
x-ray diffraction (SR-XRD) and conventional XRD. SR-XRD was
performed at the Rossendorf beamline (BM20) at the ESRF with an
x-ray wavelength of 0.154 nm. 2$\theta$-$\theta$ scans were used
to identify crystalline precipitates.

Conversion electron M\"ossbauer spectroscopy (CEMS) in
constant-acceleration mode at room temperature (RT) was used to
investigate the Fe lattice sites, electronic configuration and
corresponding magnetic hyperfine fields. The spectra were
evaluated with Lorentzian lines using a least squares fit
\cite{brand87}. All isomer shifts are given with respect to
$\alpha$-Fe at RT.

The magnetic properties were measured with a superconducting
quantum interference device (SQUID, Quantum Design MPMS)
magnetometer in the temperature range of 5-350 K. The samples were
measured with the field aligned either along the in- or
out-of-plane direction. The temperature dependence of the
magnetization was studied at a constant field and the field
dependence at a constant temperature. By magnetic measurement,
virgin ZnO is found to be purely diamagnetic with a susceptibility
of -2.65$\times$10$^{-7}$ emu/Oe$\cdot$g. This background was
subtracted from the magnetic data. To measure the temperature
dependent magnetization after zero field cooling and field cooling
(ZFC/FC), the sample was cooled in zero field from above room
temperature to 5 K. Then a 50 Oe field was applied. The ZFC curve
was measured with increasing temperature from 5 to 300 (or 350) K,
after which the FC curve was measured in the same field from 300
(or 350) to 5 K with decreasing temperature.

\section{Results} \label{section:results}
In this section, we present experimental data on structural and
magnetic properties of $^{57}$Fe implanted ZnO. Of interest in
this study are the ion-implantation induced lattice damage, the
distribution of Fe, the formation of metallic Fe nanocrystals, the
charge state of Fe, the magnetic properties, and the structure and
magnetism evolution upon post annealing. The difference between
ZnO bulk crystals and epitaxial films upon Fe implantation is also
compared.

\subsection{Lattice damage accumulation}

\subsubsection{Fluence dependence}

\begin{figure} \center
\includegraphics[scale=0.70]{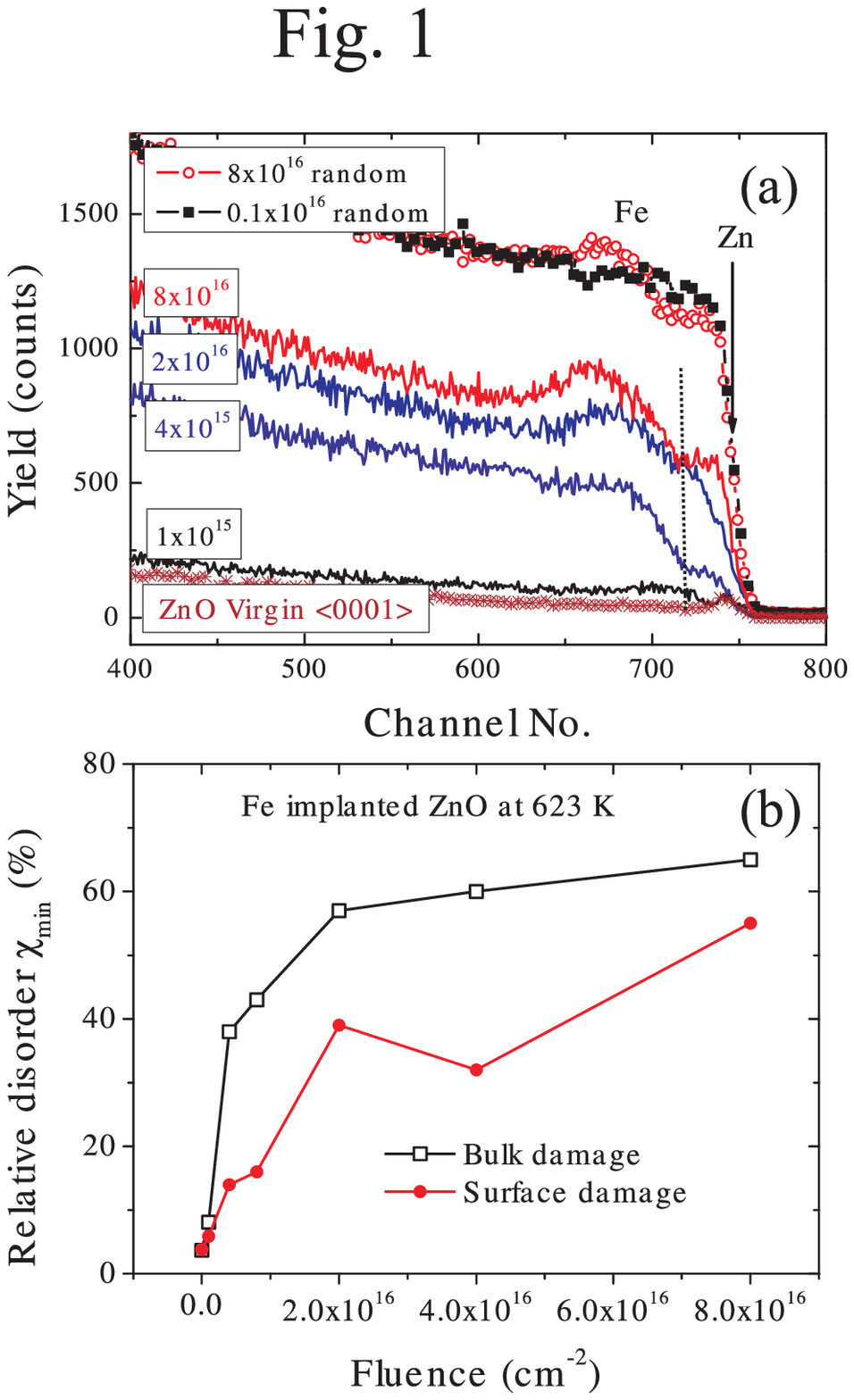}
\caption{(a) Representative RBS random and channeling spectra of
Fe implanted ZnO with the implantation energy of 180 keV. The
fluence is indicated on the channeling spectra. The dashed line
separates the damage regions of surface and bulk, where the number
of displaced atoms is maximum. (b) The ion fluence dependence of
the maximum relative disorder of the Zn lattice (\chim) at
different depth (surface and bulk).}\label{fig:RBS_FeZnO_fluence}
\end{figure}

Figure 1(a) shows representative RBS/C spectra for different Fe
fluences implanted at 623 K. The arrow labelled Zn indicates the
energy for backscattering from surface Zn atoms. The implanted Fe
ions cannot be detected for the lowest fluence
(0.1$\times$10$^{16}$ cm$^{-2}$). However they are more pronounced
as a hump in the random spectrum for a high fluence of
8$\times$10$^{16}$ cm$^{-2}$. The channeling spectrum of a virgin
sample is provided for comparison. The yield increase in the
channeling spectra mainly originates from the lattice damage due
to implantation. However, in the higher fluence case, the Fe ions
also significantly increase the RBS yields. Two features are
observed in the RBS/C spectra. One is the bimodal
\cite{kucheyev03} distribution of maximum damage depths, \ie~ in
the bulk and at the surface, separated by the dashed line in
Figure 1(a). Similar depth profiles have already been discussed by
Kucheyev \etal~\cite{kucheyev03}. In the bulk damage region the
nuclear energy-loss profile is maximum, which induces a large
number of atomic displacements. The surface damage peak is often a
sink for ion implantation induced point defects \cite{kucheyev03}.

Another feature is the saturation at larger fluences. \chim, the
ration of the channeling spectrum to the random one, is calculated
in both damage regions, as shown in Figure 1(b). Above a fluence
of 2$\times$10$^{16}$ cm$^{-2}$, both damage peaks saturate. This
is due to the strong dynamic annealing effect, \ie, migration and
interaction of defects during ion implantation \cite{kucheyev03}.
This strong dynamic annealing also makes ZnO an
$\emph{irridation-hard}$ material, \ie, it still partly persists a
crystalline state after irradiation by Fe ions up to a fluence of
8$\times$10$^{16}$ cm$^{-2}$ (\chim~of 68\%).

\subsubsection{Implantation temperature dependence}

\begin{figure} \center
\includegraphics[scale=0.7]{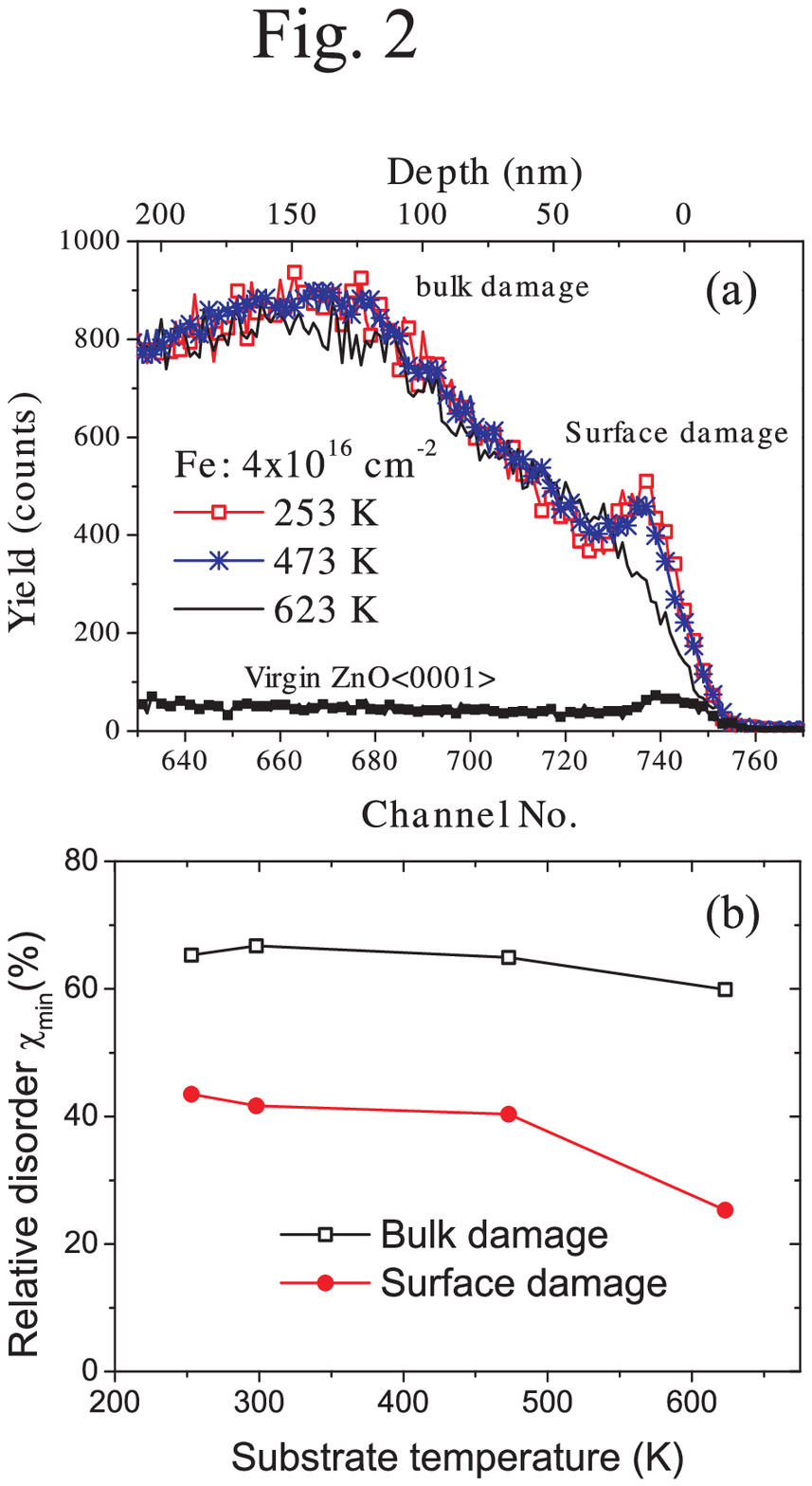}
\caption{(a) Representative RBS/C spectra with different
implantation temperature. The fluence is 4$\times$10$^{16}$
cm$^{-2}$, and implantation energy is 180 keV. (b) The calculated
\chim~for different implantation temperature, Implantation at low
temperature ($\leq$473 K) results in more damage at the surface
region.}\label{fig:RBS_FeZnO_temperature}
\end{figure}

In general increasing the substrate temperature during
implantation can suppress the lattice damage in semiconductors.
However this is not the case for ZnO. Figure 2(a) shows the channeling spectra
for Fe implanted ZnO at different implantation temperatures.
Although the surface damage peak increases drastically with
decreasing implantation temperature, the bulk damage peak is
hardly effected by implantation temperature. This can be observed
clearly in Figure 2(b). The point
defects induced by ion-beam can be significantly suppressed by
increasing the implantation temperature above 623 K. This
temperature is very critical, and below 623 K, the surface damage
peak also has no dependence on the substrate temperature. This is
very important for the electrical doping of ZnO by ion
implantation, where point defects are believe to decrease the
conductivity \cite{kucheyev02}.

\subsubsection{Recovering by post-annealing}

As shown above, the bulk damage cannot be suppressed by increasing
the implantation temperature. It has to be removed by post
annealing at higher temperature. The annealing was performed in
high vacuum in order to avoid extrinsically induced oxidation of
Fe. The temperature was varied from 823 K to 1073
K. The details have been reported in the Ref. \cite{zhou07JPD}.
Both the surface and bulk damage peaks decreased progressively
with increasing the annealing temperature and time. However even
after annealing at 1073 K for 3.5 hours, there is still
considerable damage. This is because of the high melting point of
ZnO ($\sim$2250 K). The extended defects can only be removed
completely by annealing at approximately two-thirds of the melting
temperature \cite{kucheyev01}. Therefore a high annealing
temperature (1500 K) is necessary to completely recover the
lattice structure of ZnO. However, high vacuum annealing above
1000 K also lead the decomposition of ZnO \cite{coleman:231912}.

\subsection{Fe distribution}

\begin{figure} \center
\includegraphics[scale=0.7]{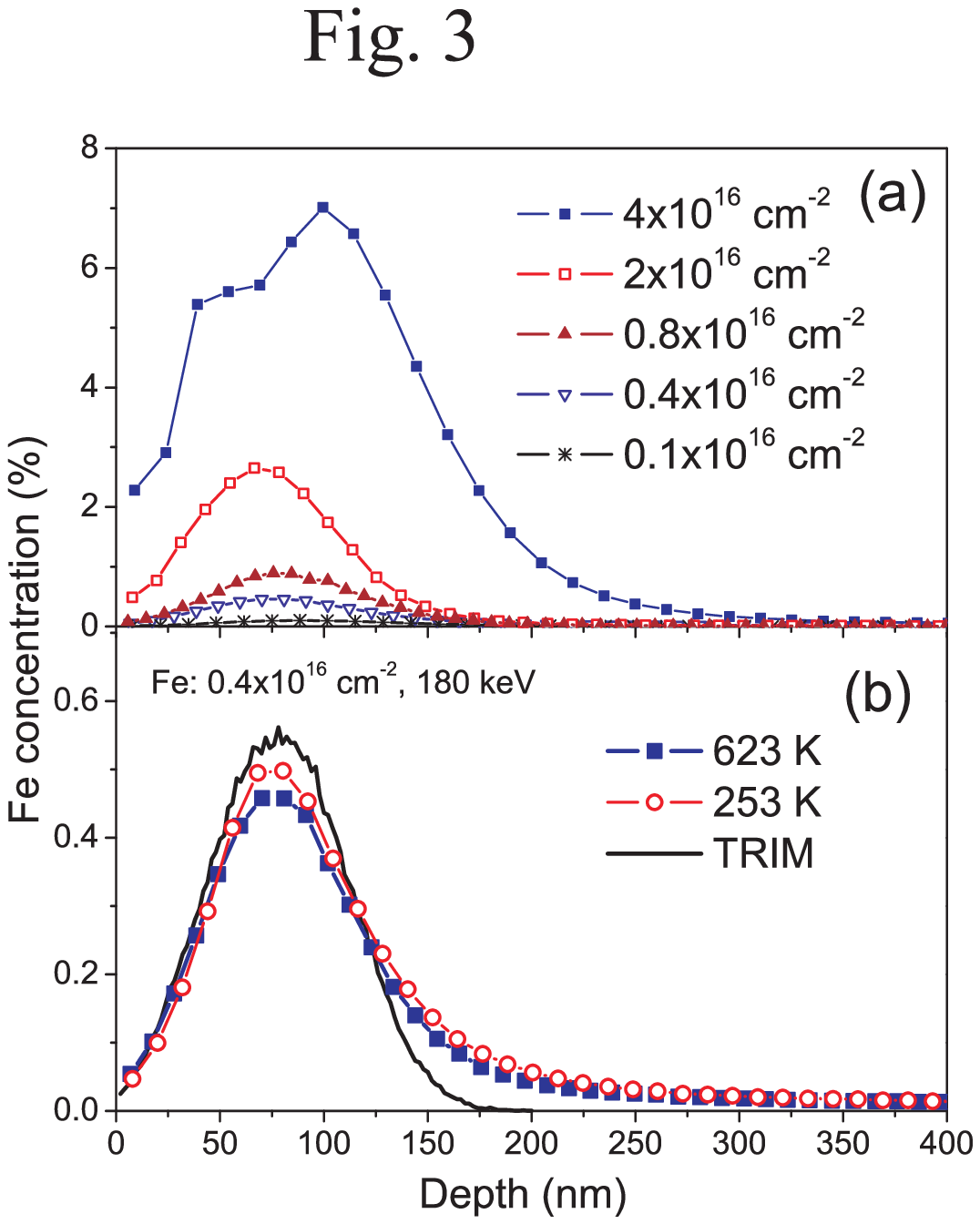}
\caption{(a) Fe distribution in the samples implanted with Fe at
623 K for different fluences measured by SIMS. The ion fluences
are indicated in the figure. (b) Distribution of Fe in ZnO
implanted at different temperatures. The TRIM \cite{trim}
simulation is presented for a comparison.}\label{fig:SIMS_FeZnO}
\end{figure}

RBS/C can give an overview of the lattice damage upon Fe
implantation. However, since the mass of Fe is smaller than Zn, it
is difficult to obtain the depth profile of the implanted Fe.
Therefore, SIMS is employed to determine the Fe depth profile (see
Figure 3(a)). It is observed that the peak
concentration of Fe increases from 0.1\% to 7\%, with a projected
range of $R_P$=(80-90)$\pm$(20-30). This is in a rather good
agreement with TRIM simulations \cite{trim}. The only discrepancy
is the high fluence sample ($4\times10^{16}$ cm$^{-2}$), where
TRIM simulations predict a peak concentration of 5\%. This is due
to the change in SIMS sensitivities for different materials, which
implies that the determined concentrations are more accurate in
the low concentration (below 1\%) regime.

In Figure 3(b) the Fe depth profile is compared
for different temperatures with the same fluence of
$0.4\times10^{16}$ cm$^{-2}$. The profile does not change
significantly due to elevating the implantation temperature from
253 K to 623 K. The slightly higher concentration for implantation
at 253 K is within the fluence error.

As discussed the in Ref. \cite{zhou07JPD}, Fe diffuses towards the
surface after high temperature annealing. The same diffusion of Fe
upon annealing was also observed by SIMS (not shown).

\subsection{Formation of Fe NCs} \label{section:Formation_Fe}

By employing SR-XRD, we have systematically investigated the
formation of Fe NCs, and its dependence on the fluence and
implantation temperature by SR-XRD.

\begin{figure} \center
\includegraphics[scale=0.9]{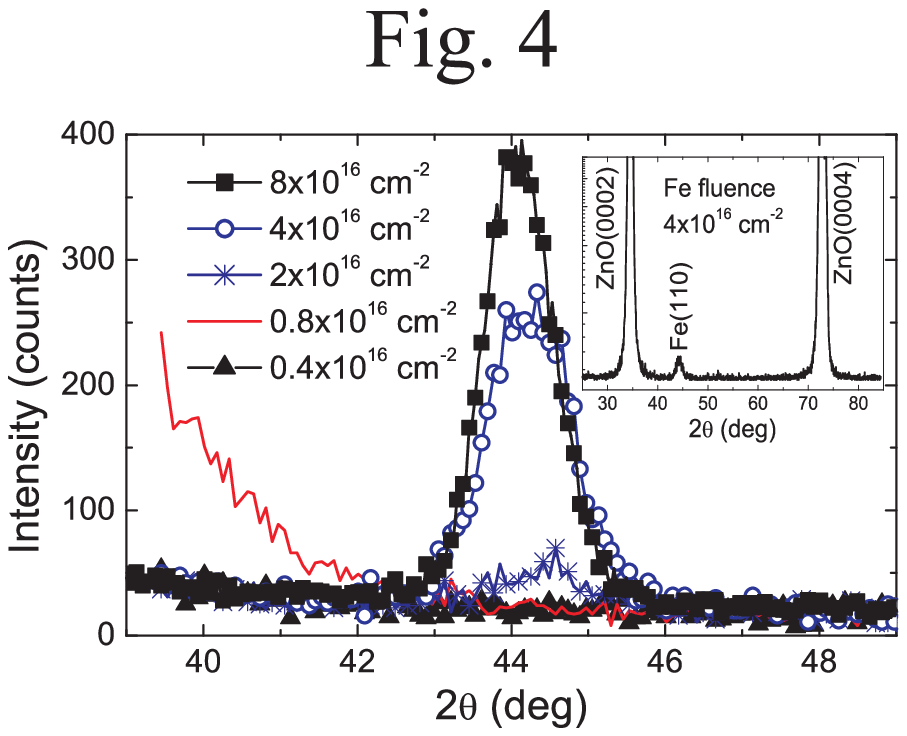}
\caption{SR-XRD 2$\theta$-$\theta$ scans of Fe implanted ZnO for
different fluences reveal the formation of crystalline Fe
nanoparticles.}\label{fig:XRD_FeZnO_fluence}
\end{figure}

\subsubsection{Fluence dependence}

\begin{table*}
\caption{\label{tab:FeZnO_sample}Structural properties of
$^{57}$Fe-implanted ZnO bulk crystals and epitaxial thin films.
The implantation energy is 180 keV.}
\begin{ruledtabular}
\begin{tabular}{ccccccc}
  Fluence & Implantation &  Peak Concentration & Peak Concentration & \multicolumn{2}{c}{\chim~(RBS/C)} & Metallic Fe \\
  (cm$^{-2}$) & Temperature (K) & (TRIM simulation) & (SIMS) & Bulk & Surface & Formation \\
  \hline
  0.1$\times$10$^{16}$ & 623 & 0.14\% & 0.1\% & 8.1\% & 5.9\% & No \\
  0.4$\times$10$^{16}$ & 623 & 0.55\% & 0.46\% & 38\% & 14\% & No \\
  0.8$\times$10$^{16}$ & 623 & 1.1\% & 0.89\% & 43\% & 16\% & No \\
  2$\times$10$^{16}$ & 623 & 2.7\% & 2.6\% & 57\% & 39\% & $\alpha$-Fe \\
  4$\times$10$^{16}$ & 623 & 5.5\% & 6.0\% & 60\% & 32\% & $\alpha$-Fe \\
  8$\times$10$^{16}$ & 623 & 11\% & - & 65\% & 55\% & $\alpha$-Fe \\
  \\
  0.4$\times$10$^{16}$ & 253 & 0.55\% & 0.5\% & 31\% & 16\% & No \\
  4$\times$10$^{16}$ & 253 & 5.5\% & 5.5\% & 65\% & 43\% & No \\
  4$\times$10$^{16}$ & 298 & 5.5\% & - & 65\% & 40\% & No \\
  4$\times$10$^{16}$ & 473 & 5.5\% & - & 66\% & 42\% & No \\
  \\
  4$\times$10$^{16} (\footnotemark[1])$ & 623 K & 5.5\% & - & 44\% & - & $\alpha$ and $\gamma$-Fe \\
\end{tabular}
\end{ruledtabular}
\footnotetext[1] {ZnO epitaxial thin films.}
\end{table*}

\begin{table*}
\caption{\label{tab:FeZnO}Structural and magnetic properties for
Fe-implanted ZnO. The ferromagnetic fraction corresponds to the
percentage of ferromagnetic Fe (at 5 K) compared with all
implanted Fe ions. The crystallite size evaluated by ZFC
magnetization is only for $\alpha$-Fe NCs.}
\begin{ruledtabular}
\begin{tabular}{cccccccccc}
  \multicolumn{3}{c}{Sample} &  Crystallite size & T$_B$ & \multicolumn{2}{c}{Crystallite size} & Saturation & Ferromagnetic & Coercivity \\
  Fluence & T$_{imp.} $ & T$_{ann.}$ &  (XRD) & (ZFC) & Eq.\ref{blocking_squid} & Eq. \ref{ZFCM} & magnetization (5 K) & Fe fraction\footnotemark[2] & at 5 K \\
  (cm$^{-2}$) & (K) & (K) &  (nm) & (K) & \multicolumn{2}{c}{(nm)} & ($\mu_B$/Fe) & (\%) & (Oe) \\
  \hline
  2$\times$10$^{16}$ & 623 & - &  5.6 & 38 & 8 & 6.6 & 0.08 & 3.6 & 600 \\
  4$\times$10$^{16}$ & 623 & - &  7.1 & 137 & 12 & 8.9 & 0.24 & 11 & 360 \\
  8$\times$10$^{16}$ & 623 & - &  8.9 & 212 & 14 & 11.3 & 0.13 & 5.9 & 360 \\
  4$\times$10$^{16}$ & 623 & 823 &  9.4 & 200 & 14 & 10.2 & 0.34 & 15 & 360 \\
  \\
  4$\times$10$^{16}$ & 253 & 823 & - & 295 & 16 & 9.5 & 0.52 & 24 & 370 \\
  \\
  4$\times$10$^{16}$\footnotemark[1] & 623 & - & -/6 ($\alpha$/$\gamma$-Fe) & 26 & 7 & 4.6 & 0.55 & 25 & 220  \\
  4$\times$10$^{16}$\footnotemark[1] & 623 & 823 & 8.1/11 ($\alpha$/$\gamma$-Fe) & 280 & 15 & 10.2 & 1.3 & 59 & 220 \\
\end{tabular}
\end{ruledtabular}
\footnotetext[1] {ZnO epitaxial layers implanted with Fe at 180
keV and 623 K. }\footnotetext[2] {Calculated by comparing the
saturation magnetization with the value (2.2 $\mu_B$/Fe) for bulk
Fe. }
\end{table*}

Figure 4 shows the SR-XRD pattern (focused on Fe(110) peak) as a
function of fluence. At a low fluence (0.1$\times$10$^{16}$ to
0.8$\times$10$^{16}$ cm$^{-2}$), no crystalline Fe NCs could be
detected, while above a fluence of 2$\times$10$^{16}$ cm$^{-2}$),
an Fe(110) peak appears and increases with fluence. The inset
shows a wide range scan for the high fluence sample
(4$\times$10$^{16}$ cm$^{-2}$). No other Fe-oxide (Fe$_2$O$_3$,
Fe$_3$O$_4$, and ZnFe$_2$O$_4$) particles are detected in the
as-implanted state. The full width at half maximum (FWHM) of the
Fe(110) peak decreases with fluence, indicating a growth of the
average diameter of these NCs. The crystallite size is calculated
using the Scherrer formula \cite{scherrer}.

\begin{equation}\label{scherrer}
    d=0.9\lambda/(\beta\cdot\cos\theta)
\end{equation} where $\lambda$ is the wavelength of the x-ray, $\theta$ the Bragg angle, and $\beta$ the FWHM of
2$\theta$ in radians. The crystallite size was estimated using Eq.
\ref{scherrer} and is listed in table \ref{tab:FeZnO}.

Note, that only one peak of Fe(110) appears in the inset of Figure
4. This indicates a texture of the Fe NCs. However no texture
behavior is found even for the highest fluence sample in pole
figure measurements on Fe(110) and Fe(200) (not shown). This could
be due to the difference in the crystalline symmetry of hexagonal
ZnO (six fold symmetry) and bcc-Fe (four fold symmetry). For a
bcc-crystal, one cannot find a six-fold symmetry viewed from any
direction. In contrast, hcp-Co(0001) and fcc-Ni(111) NCs, which
are six-fold symmetric, are found to be crystallographically
oriented inside ZnO matrix. This highly ordered orientation allows
them to be detected even by laboratory XRD
\cite{zhou06}\cite{zhou07CoNi}.

\subsubsection{Implantation temperature dependence}

\begin{figure} \center
\includegraphics[scale=0.7]{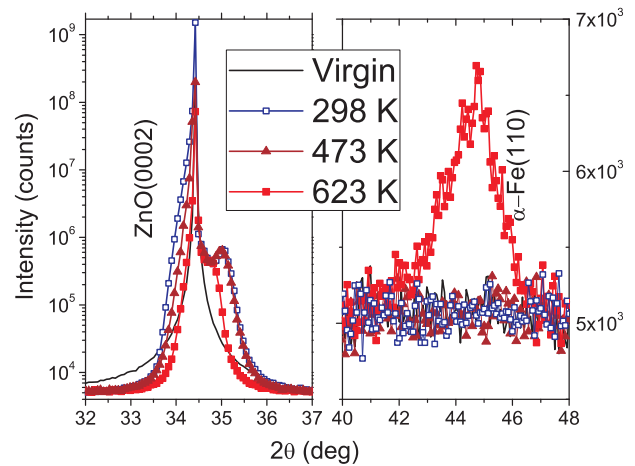}
\caption{SR-XRD 2$\theta$-$\theta$ scans of Fe implanted ZnO with the same fluence of 4$\times$10$^{16}$ cm$^{-2}$ at 
parts. Only the sample implanted at 623 K shows $\alpha$-Fe
precipitates.}\label{fig:XRD_FeZnO_temperature}
\end{figure}

SR-XRD was also performed for the samples with an Fe fluence of
4$\times$10$^{16}$ cm$^{-2}$ implanted at different temperatures
from 253 K to 623 K. As shown in Figure 5, for implantation
temperatures of 473 K and below, no crystalline Fe could be
detected. This is also confirmed by the CEMS results (shown
later), where the Fe$^0$ state appears only at an implantation
temperature of 623 K. Note the asymmetry of the ZnO(0002)
diffraction peaks in Figure 5. Shoulders on the right side
(smaller lattice constant) are clearly observed. These shoulders
decrease with increasing implantation temperature, and can
therefore be associated with lattice damage or ZnO substituted
with Fe. In view of a detailed study of ion implantation into GaN
where the implantation induces a lattice expansion of GaN (a
shoulder at left side)
\cite{GaN_expension}\cite{GaN_expension_review}, we rather
attribute the observed right side shoulders to ZnO substituted
with Fe. In the 623 K implantation, metallic Fe NCs start to form,
therefore the substitution is reduced.

\subsubsection{Growth with post-annealing}

After thermal annealing at 823 K for 15 min, more $\alpha$-Fe NCs
of a larger size are formed for the samples implanted at 623 K.
After 1073 K and 15 min annealing, the $\alpha$-Fe almost
disappears and ZnFe$_2$O$_4$ starts to form. The details of the
structure and magnetism evolution upon thermal annealing can be
found in the Ref. \cite{zhou07JPD}.

\subsection{Charge state of Fe} \label{section:charge_state}

CEMS allows one to identify different site occupations, charge and
magnetic states of $^{57}$Fe. The hyperfine parameters calculated
according to the evaluations of the spectra are given in Table
\ref{tab:hyperfine_FeZnO}. All isomer shifts are given relative to
an $\alpha$-Fe reference foil. In general, the implanted Fe occupy
three different states: metallic Fe, Fe$^{2+}$ and Fe$^{3+}$ ions
dispersed in the ZnO matrix and finally Fe$^{3+}$ in Zn-ferrites.
The hyperfine interaction parameters obtained from the best fits
are different from that of ferromagnetic $\alpha$ or
$\gamma$-Fe$_2$O$_3$ or ferrimagnetic Fe$_3$O$_4$. Hence, the
presence of these phases was excluded. The dispersed ionic Fe
could substitute onto Zn site.

\begin{table*}
\caption{\label{tab:hyperfine_FeZnO}Hyperfine parameters obtained
from the evaluation of CEMS for samples implanted or annealed at
different temperatures. The fluence was 0.4$\times$10$^{16}$
cm$^{-2}$ for the first sample, while 4$\times$10$^{16}$ cm$^{-2}$
for all other samples. The codes of S1, S2 and S3 notate the
samples for post-annealing process. The notations for the fitting
lines are given as S (singlet), D (doublet) and M (sextet).}
\begin{ruledtabular}
\begin{tabular}{ccc|cc|ccc|ccc|ccc}
  \multicolumn{3}{c} {Sample} & \multicolumn{2}{c} {S} & \multicolumn{3}{c} {D(I)} & \multicolumn{3}{c} {D(II)} & \multicolumn{3}{c} {M ($\alpha$-Fe)}\\
  Code & T$_{imp.}$ & T$_{ann.}$ & FR\footnotemark[1] & IS\footnotemark[2] & FR\footnotemark[1] & IS\footnotemark[2] & QS\footnotemark[3] & FR\footnotemark[1] & IS\footnotemark[2] & QS\footnotemark[3] & FR\footnotemark[1] & IS\footnotemark[2] & B$_{hf}\footnotemark[4]$  \\
& (K) & (K) & (\%) & ($mm/s$) & (\%) & ($mm/s$) & ($mm/s$) & (\%) & ($mm/s$) & ($mm/s$) & (\%) & ($mm/s$) & (T) \\
  \hline
  & 623 & - & 27.7 & 0.57 & 13.7 & 0.31 & 0.75 & 58.6 & 0.81 & 0.79 & - & - & - \\
  \\
  S1 & 623 & - & 32.8 & 0.53 & 31.5 & 0.78 & 1.29 & 23.2 & 0.96 & 0.58 & 12.5 & 0.06 & 30.5 \\
  S1 & 623 & 823  & 42.6 & 0.42 & 16.7 & 0.68 & 1.52 & 22.5 & 0.94 & 0.54 & 18.2 & 0.07 & 31.7 \\
  S1 & 623 & 1073  & - & - & 100 & 0.35 & 0.43 & - & - & - & - & - & - \\
  \\
  S2 & 253 & - & 13.6 & 0.22 & 14.1 & 0.24 & 0.65 & 72.3 & 0.92 & 0.97 & - & - & - \\
   & 473 & - & 22.2 & 0.32 & 9.7 & 0.27 & 0.63 & 68.1 & 0.94 & 0.77 & - & - & - \\
  S2 & 253 & 823 & 6.2 & -0.09 & 46.8 & 0.42 & 0.39 & 18.2 & 0.88 & 0.51 & 28.8 & 0.04 & Dist.\footnotemark[5] \\
  \\
  S3\footnotemark[6] & 623 & - & 23.0 & -0.09 & 26.3 & 0.45 & 0.32 & 37.3 & 0.91 & 0.80 & 13.4 & 0 & Dist.\footnotemark[5] \\
  S3 & 623 & 823 & 13.9 & -0.09 & 31.4 & 0.45 & 0.35 & 15.9 & 0.92 & 0.58 & 38.8 & 0.02 & Dist.\footnotemark[5] \\
\end{tabular}
\end{ruledtabular}
\footnotetext[1]{Fraction corresponding to the relative area of
the subspectrum.} \footnotetext[2]{Isomer shift: 0 mm/s for
$\alpha$-Fe, 0.7-1.2 mm/s for Fe$^{2+}$, 0.2-0.7 mm/s for
Fe$^{3+}$, and -0.1 mm/s for $\gamma$-Fe.}
\footnotetext[3]{Quadrupole splitting.} \footnotetext[4]{Magnetic
hyperfine field.} \footnotetext[5]{Hyperfine field
distribution.}\footnotetext[6]{ZnO epitaxial thin films.}
\end{table*}

\subsubsection{Fluence dependence}

Figure 6(a) and 6(b) show the comparison of Fe implanted ZnO at
623 K with a fluence of 0.4$\times$10$^{16}$ and
4$\times$10$^{16}$ cm$^{-2}$, respectively. In spectrum (a), the
singlet S and doublet D(I) are attributed to Fe$^{3+}$, while the
doublet D(II) is from Fe$^{2+}$. In the high fluence sample
(spectrum (b)), the majority of Fe are ionic states Fe$^{3+}$
(singlet S) and Fe$^{2+}$ (doublet D(I) and D(II)), while a
considerable fraction of a sextet associated to $\alpha$-Fe is
present (sextet M). The formation of $\alpha$-Fe is in agreement
with SR-XRD observation (Figure 4). At room temperature, all
Fe$^{2+}$ and Fe$^{3+}$ show no ferromagnetic interaction. Later
on in subsection \ref{section:magnetic_properties}, we show that
even at 5 K the measured ferromagnetism can only be attributed to
$\alpha$-Fe NCs.

\begin{figure}\center
\includegraphics[scale=0.4]{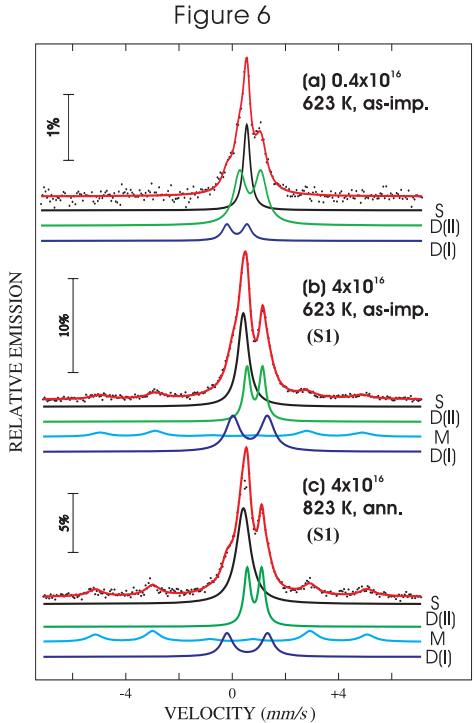}
\caption{Room temperature CEMS for ZnO bulk crystals implanted
with $^{57}$Fe with different fluences, and with post-annealing.
The notations for the fitting lines are given as S (singlet), D
(doublet) and M (sextet). The fluence and the process temperatures
are indicated.}\label{fig:CEMS_fluence_ZnO}
\end{figure}

\subsubsection{Implantation temperature dependence}

Figure 7(a) and 7(b) shows CEMS for the samples implanted at low
temperatures, 473 K and 253 K, respectively, with a $^{57}$Fe
fluence of 4$\times$10$^{16}$ cm$^{-2}$. In these two samples,
ionic Fe are the dominant charge states: Fe$^{3+}$ (S and D(I)),
and Fe$^{2+}$ (D(II)). In contrast to Figure 6(b), there is no
detectable $\alpha$-Fe in these two samples. This is also in
agreement with SR-XRD results (Figure 5), where up to an
implantation temperature of 473 K no $\alpha$-Fe is found.

\begin{figure} \center
\includegraphics[scale=0.45]{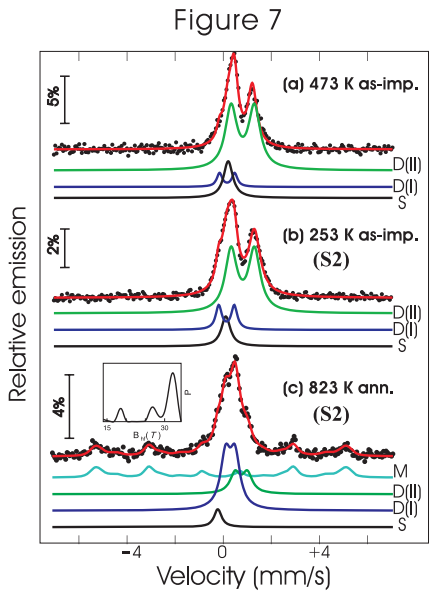}
\caption{Room temperature CEMS for ZnO implanted with $^{57}$Fe at
253 K and subsequent annealed at 823 K for 15 min. The notations
for the fitting lines are given as S (singlet), D (doublet) and M
(sextet). On the right side of the spectra, the probability
distribution P for the magnetic hyperfine field (B$_{hf}$, solid
lines)are given.}\label{fig:CEMS_lowT_ZnO}
\end{figure}
\subsection{Implantation Energy dependence}

\subsubsection{Evolution with post-annealing}
The post-annealing was performed on selected samples: S1 and S2
(Table \ref{tab:hyperfine_FeZnO}). They were implanted with the
same fluence of 4$\times$10$^{16}$ cm$^{-2}$ at 623 K and 253 K,
respectively. For sample S1, upon annealing at 823 K for 15 min,
the intensity of the sextet increases up to 18.2\% while the
fraction of Fe$^{2+}$ (doublet D(I)) decreases, suggesting the
growth of the $\alpha$-Fe nanoparticles and the recovery of
lattice defects (Figure 6(c)). Moreover, the value for the
magnetic hyperfine field B$_{hf}$ increases upon annealing and
moves closer (from 30.5 T to 31.7 T) to the known value for bulk
$\alpha$-Fe (33 T). For sample S2, after annealing at 823 K for 15
min, the relative fraction of metallic $\alpha$-Fe increases up to
28.8\%. The hyperfine field is distributed with maxima at 18 T, 27
T and mostly at 32.5 T ($\alpha$-Fe). Comparing with the annealing
of sample S1, a larger fraction of $\alpha$-Fe is formed in the
annealed sample S2. This is consistent with a larger magnetization
measured by SQUID (shown later). In addition, a small fraction of
singlet (S) presents, which is attributed to $\gamma$-Fe according
to the isomer shift.

Higher temperature (1073 K) annealing was performed on sample S1
and has been reported in Ref \cite{zhou07JPD}. After annealing at
1073 K for 3.5 hours, Fe$^{3+}$ is the only charge state, and
Zn-ferrites (ZnFe$_2$O$_4$) are formed and are
crystallographically oriented inside ZnO matrix.

\subsection{Magnetic properties Fe implanted ZnO}\label{section:magnetic_properties}

In the previous sections, we have reported a thorough
investigation on the structural properties, and the charge states
of Fe. The main conclusion can be summarized as follows (i) upon
implantation at a temperature of 623 K, a small part (around 12\%)
of the implanted Fe ions forms as crystalline Fe already in the
as-implanted state, while the major part of the implanted Fe is in
ionic states (Fe$^{2+}$, and Fe$^{3+}$); (ii) implantation at a
low temperature (253 K) suppresses the metallic Fe formation, and
the implanted Fe ions are in ionic states, but they are not
magnetically coupled at room temperature; (iii) post-annealing at
823 K largely enhances the Fe NC formation in all implanted
samples for both implantation temperatures (253 K and 623 K).
Since CEMS was performed at room temperature only, the magnetic
properties of metallic and ionic Fe at low temperature could not
be determined. Here we present the results from SQUID magnetometry
measured from 5 K to 350 K. We will show that the metallic Fe NCs
are superparamagnetic, and they are the predominant contribution
to the measured ferromagnetic response even at 5 K. In contrast
the ionic Fe is not ferromagnetically coupled even at 5 K.

\subsubsection{Superparamagnetism of Fe NCs}

For magnetic nanoparticles, the formation of domain walls is
energetically unfavorable and below a certain size (typically in
the range of 15 to 30 nm depending on the material), the particle
stays in a single-domain configuration. The magnetism of a single
nanoparticle in a solid matrix can be described by the N\'eel
process \cite{respaud}. If the particle size is sufficiently
small, above a particular temperature (so-called blocking
temperature, T$_B$) thermal fluctuations dominate and no preferred
magnetization direction can be defined. Such a system of
superparamagnetic particles does not exhibit a hysteresis loop
above T$_B$; therefore the coercivity (H$_C$) and the remanence
(M$_R$) are both zero. Phenomenologically there are two
characteristic features in the temperature dependent magnetization
of a nanoparticle system. One is the irreversibility of the
magnetization in a small applied field (\eg~50 Oe) after zero
field cooling and field cooling (ZFC/FC) \cite{respaud}. The other
is the drastic drop of the coercivity and of the remanence at a
temperature close to or above T$_B$ \cite{shinde04}.

For a dc magnetization measurement in a small magnetic field by
SQUID, the blocking temperatue T$_B$ is given by
\begin{equation}\label{blocking_squid}
    T_{B}\approx\frac{K_{eff}V}{30k_B}
\end{equation} where $K_{eff}(V)$ is the anisotropy energy density, $V$ the particle volume, and $k_B$ the Boltzmann constant \cite{respaud}. With this equation, one can estimate the particle size \cite{shinde04}. However, in any fine particle system, there is a distribution of
particle sizes, which is usually assumed as a log-normal
distribution $D(V)$.
\begin{equation}\label{size_distribution}
    D(V)=\frac{A}{\sqrt{2\pi}\sigma_LV}exp[-\frac{ln[V/V_{mean}]}{2{\sigma_L}^2}]
\end{equation} where $V_{mean}$ is the most probable value, and
$\sigma_L$ is the standard deviation.

Such a volume distribution results in a distribution of blocking
temperatures $T_B(V)$. The ZFC magnetization can be calculated as
follows \cite{respaud}.
\begin{widetext}
\begin{equation}\label{ZFCM}
    M_{ZFC}(B,T)=\frac{{M_s}^2(T)B}{3k_BT}\frac{1}{N}\int_0^{V_{limit}(T)}{V^2D(V)}dV+\frac{{M_s}^2(T)B}{3K_{eff}}\frac{1}{N}\int_{V_{limit}(T)}^\infty{VD(V)}dV
\end{equation}
\end{widetext}
where $M_s$ is the spontaneous magnetization of the particle,
$D(V)$ the volume distribution, $V_{limit}(T)$
($=30k_BT$/$K_{eff}$) the maximum volume in the superparamagnetic
state, N the normalizing factor, and $k_B$ the Boltzmann constant.
$M_s$ is assumed to be a constant independent of temperature
\cite{respaud}\cite{farle}. The first integral represents the
contribution of the superparamagnetic particles, while the second
corresponds to the blocked ones. A more precise determination of
the size should be performed by fitting the ZFC curve with the
equation \ref{ZFCM}. Figure 8 shows the fitting on the ZFC
magnetization curve, and the corresponding size distribution. In
the fit, $K_{eff}(V)$=5$\times$10$^4$ J$m^{-3}$ is treated as a
constant.

\begin{figure} \center
\includegraphics[scale=0.8]{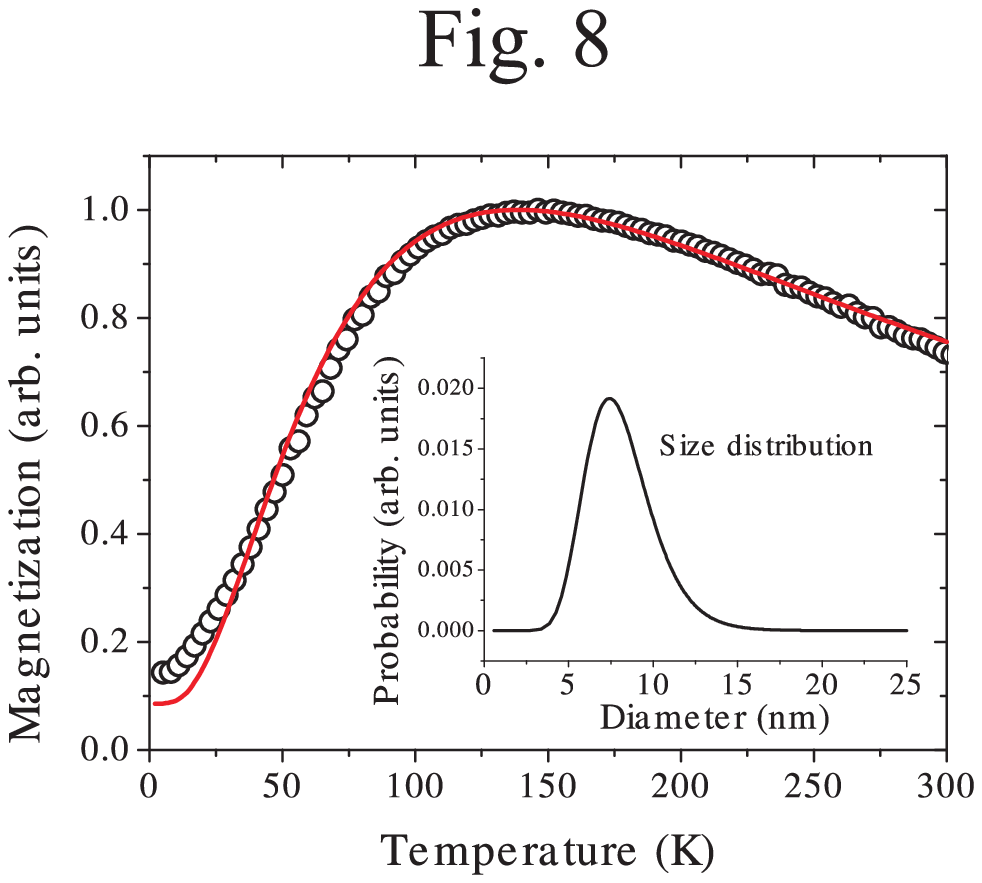}
\caption{ZFC magnetization vs temperature for the as-implanted
sample with the fluence of 4$\times$10$^{16}$ cm$^{-2}$. Solid
line is the best fitting using the Eq. \ref{ZFCM}. Inset shows the
size distribution of the Fe NCs deduced from the analysis of ZFC
magnetization.}\label{fig:simulation111}
\end{figure}

\subsubsection{Fluence dependence}

\begin{figure} \center
\includegraphics[scale=0.8]{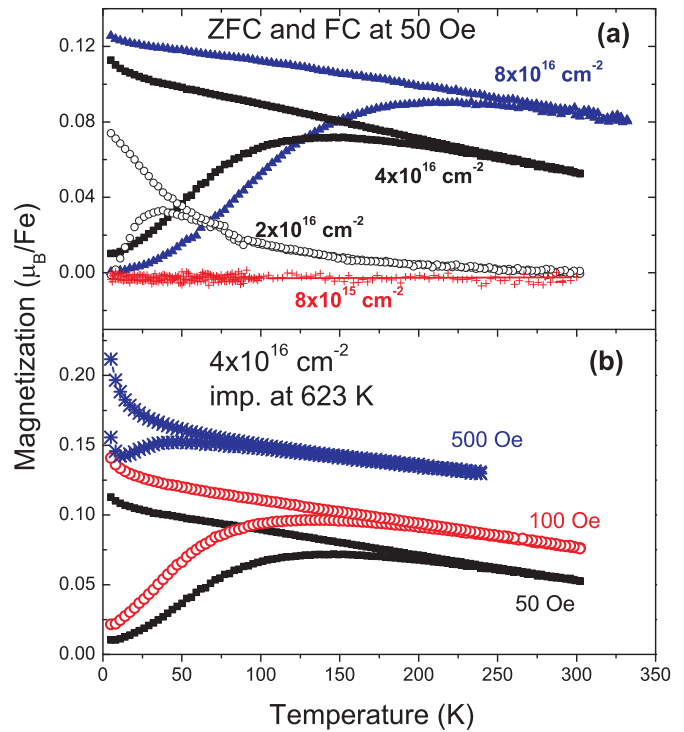}
\caption{(a) Magnetization curves with an applied field of 50 Oe
after ZFC/FC for the Fe implanted ZnO. With increasing fluence,
the Fe NCs are growing in size, resulting in a higher blocking
temperature; (b) ZFC/FC curves with different applied fields. The
blocking temperature decreases progressively with increasing
field.}\label{fig:ZFCFC_FeZnO_fluence}
\end{figure}

Figure 9(a) shows the ZFC/FC magnetization curves in a 50 Oe field
for different fluences of Fe implanted ZnO. The FC curves for low
fluences of 0.1$\times$10$^{16}$ (not shown for clarity) and
0.8$\times$10$^{16}$ cm$^{-2}$ completely overlap with the
corresponding ZFC curves at values close to zero. No
superparamagnetic particles are present in the two samples. For
larger fluences (above 2$\times$10$^{16}$ cm$^{-2}$), a distinct
difference in ZFC/FC curves was observed. ZFC curves show a
gradual increase (deblocking) at low temperatures, and reach a
broad peak with a maximum, while FC curves continue to increase
with decreasing temperature. The broad peak in the ZFC curves is
due to the size distribution of Fe NCs. In this paper, the
temperature at the maximum of the ZFC curve is taken as the
average blocking temperature (later referred as T$_B$). At a much
higher temperature than T$_B$, FC curves still depart from
corresponding ZFC curves, which distinguish the Fe particle system
from a conventional spin-glass system where the FC curve merges
together with ZFC curve just at T$_B$ and shows a plateau below
T$_B$ \cite{PhysRevLett.91.167206}. The ZFC/FC curves are general
characteristics of magnetic nanoparticle systems with a broad size
distribution \cite{tsoi:014445}. T$_B$ increases with the fluence,
\ie~the size of nanoparticles. Table \ref{tab:FeZnO}~lists the
average size of Fe NCs calculated by Eq. \ref{blocking_squid} and
by XRD data (Eq. \ref{scherrer}), and simulated by Eq. \ref{ZFCM}.
Although the trend is the same for all calculations, the values
from Eq.\ref{blocking_squid} are larger than that from Eq.
\ref{ZFCM} and from XRD data. Given the large size distribution in
the present magnetic nanoparticle system, T$_B$ is overestimated
by taking the temperature at the maximum of the ZFC curve
\cite{farle}. This explains why Eq. \ref{blocking_squid} gives a
large average particle size. For the fitting according to Eq.
\ref{ZFCM}, one has to note that $M_s$ and $K_{eff}(V)$ are
assumed to be temperature independent, and the interaction between
the NCs is ignored. These effect contribute to the error bars
\cite{jacobsohn:321}. Nevertheless, from both techniques (XRD, and
ZFC magnetization), we have determined the size of Fe NCs and its
distribution.

Figure 9(b) shows ZFC/FC curves with
an applied field ranging from 50 Oe to 500 Oe for the sample with
fluence of 4$\times$10$^{16}$ cm$^{-2}$. T$_B$ in the ZFC curves
clearly shifts to lower temperatures with increasing field. This
behavior is expected for magnetic nanoparticles since the reduced
energy barrier caused by the external magnetic field allows the
reorientation of the superparamagnetic moments by thermal
fluctuations at lower temperatures \cite{park04}\cite{shinde04}.

\begin{figure} \center
\includegraphics[scale=0.8]{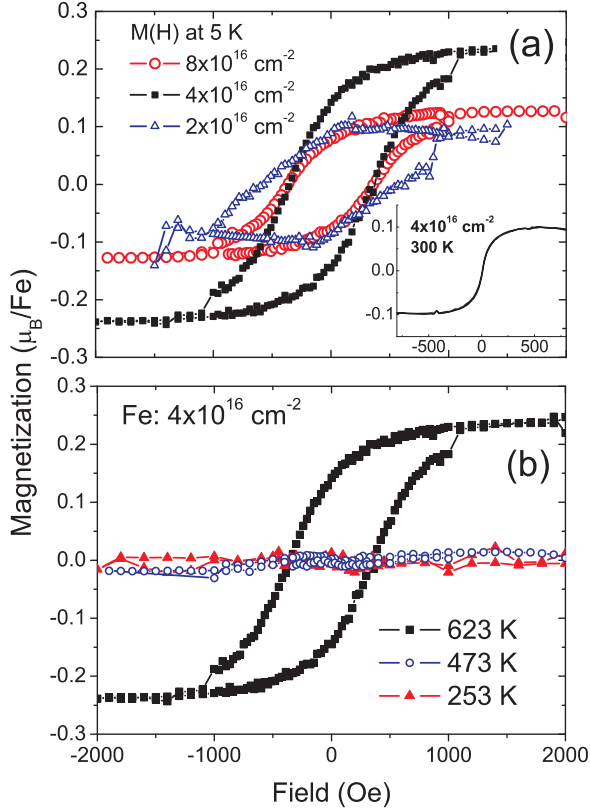}
\caption{(a) Hysteresis loops measured at 5 K for Fe implanted ZnO
with different fluence. The inset shows the room temperature M-H
loop for the sample implanted with a fluence of 4$\times$10$^{16}$
cm$^{-2}$. (b) Hysteresis loops measured at 5 K for Fe implanted
ZnO at different implantation temperatures with the same fluence
of 4$\times$10$^{16}$
cm$^{-2}$.}\label{fig:MH_FeZnO_fluence_temperature}
\end{figure}

Figure 10(a) shows the magnetization versus field reversal (M-H)
of samples implanted with large fluences. At 5 K, hysteretic
behaviors were observed for all three samples. The saturation
moment is increased with increasing fluence, however the
coercivity is decreased from 600 Oe for a fluence of
2$\times$10$^{16}$ cm$^{-2}$ to 330 Oe for larger fluences (see
table I). This can be explained by the enhanced coercivity effect
for the interfacial spins, which increases with decreasing the
size of nanoparticles \cite{Battle}. The inset shows the M-H curve
at 300 K for the sample implanted with the fluence of
4$\times$10$^{16}$ cm$^{-2}$. As expected for a magnetic
nanoparticle system, above the blocking temperature, both
remanence and coercivity drop to zero.

\subsubsection{Implantation temperature dependence}

Figure 10(b) shows the
magnetization versus field reversal of samples implanted with Fe
(4$\times$10$^{16}$ cm$^{-2}$) at different implantation
temperatures. Only the sample implanted at 623 K shows a
hysteretic behavior due to the presence of Fe NCs, while the other
samples implanted at 473 K or below show no ferromagnetic response
down to 5 K. This is in full agreement with SR-XRD and CEMS
measurements.

\subsubsection{Post annealing effect}

The magnetic properties of the samples implanted at 623 K have
been reported in a previous paper \cite{zhou07JPD}. The main
conclusions are the following: the annealing at 823 K results in
the growth of $\alpha$-Fe NCs. During annealing at 1073 K the
majority of the metallic Fe is oxidized; after a long term
annealing at 1073 K, crystallographically oriented ZnFe$_2$O$_4$
NCs form. Here we mainly present the annealing at 823 K for the
samples implanted at 253 K. Due to the different initial state
from the 623 K implanted samples, the same annealing temperature
leads to different results.

\begin{figure} \center
\includegraphics[scale=0.8]{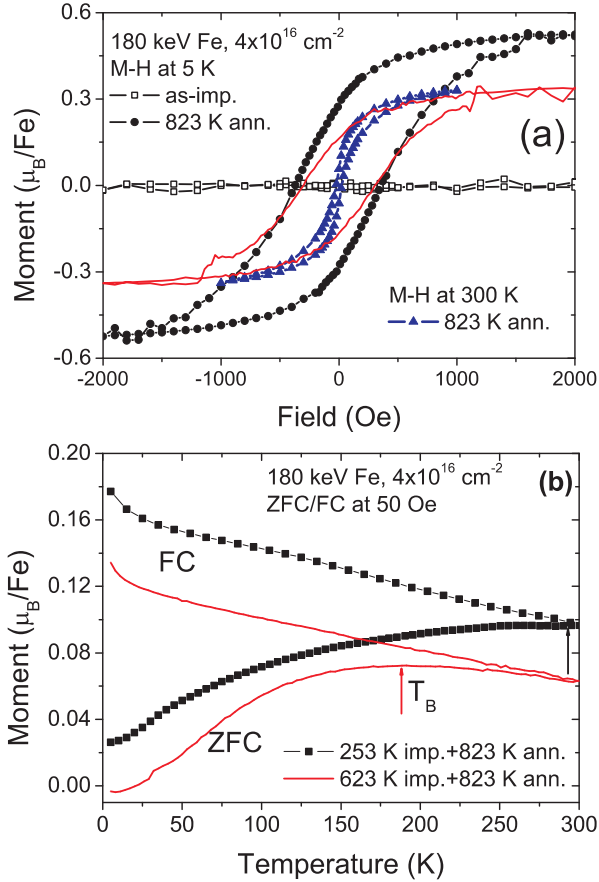}
\caption{Hysteresis loops measured at 5 K for ZnO implanted with
180 keV Fe at 253 K (as-implanted and post annealed). The fluence
is 4$\times$10$^{16}$ cm$^{-2}$. For comparison, the sample
implanted with the same energy and fluence, but at a high
implantation temperature (623 K), after the same annealing process
is shown as the solid line. (b) ZFC/FC magnetization of the sample
after 823 K annealing. The arrows indicate the blocking
temperatures.}\label{fig:SQUID_FeZnO_180keV_lowTann}
\end{figure}

Figure 11 shows the magnetic properties of the samples implanted
at 253 K with subsequent post annealing. In the as-implanted
state, there is no ferromagnetism down to 5 K, while after 823 K
annealing, magnetization of 0.52 $m_B$/Fe was observed. The ZFC/FC
magnetization curves show the characteristics of magnetic
nanoparticle system. According to SR-XRD and CEMS results, we
attribute this to Fe NCs. The ZFC curve is very broad and T$_B$ is
above room temperature. The M-H curves at 300 K for both cases are
still open, although with much smaller coercivity and remanence
compared with 5 K. However the magnetic properties are quite
different from the 623 K implanted sample after post annealing at
823 K, where the T$_B$ is well below 300 K, and at 300 K there is
neither coercivity nor remanence. We will discuss this difference
in section \ref{section:discussions}.

\subsubsection{Magnetic anisotropy of Fe NCs}

M-H loops were also measured for selective samples which have been
implanted with a fluence of 4$\times$10$^{16}$ cm$^{-2}$ with the
field applied perpendicular to the sample surface. Figure 12 shows
the comparison of the magnetization between the in-plane and
out-of-plane direction at 5 K. The in-plane (parallel to the ZnO
surface) is the easy axis, while the out-of-plane (perpendicular)
is the hard axis. At 5 K, the coercivity of the easy axis is
around 360 Oe, and the ratio of M$_R$/M$_S$ (remanence and
saturation moment) is around 58\%. The anisotropy energy, $K$, can
be calculated according to the equation of $K$=$M_AH_s/2$, where
$M_s$ is the saturation moment of 4$\pi$$M_s$=22000 G, $H_A$ is
the effective anisotropy field. Indeed $H_A$ is rather difficult
to be deduced since it is not easy to measure a \emph{real}~hard
axis loop with SQUID magnetometry without a precise control of the
sample alignment. Moreover the size distribution of Fe NCs could
result in a distribution of $H_A$. Therefore, we deduce a lower
and upper limit of $H_A$ according to the shape of the hard axis
loop. Using this approach, the anisotropy energy is estimated to
be in the range of (1.8-3.3)$\times$10$^{5}$ Jm$^{-3}$. It is
larger than the magnetocrystalline anisotropy, and around one
order of magnitude larger than the uniaxial anisotropy observed in
Fe thin films \cite{shaw:094417} and micro-scale Fe nanomagnets
\cite{pulwey:7995}. If the Fe NCs are assumed to be sphere-like,
their magnetism should be isotropic, unless they are textured.
However as found by XRD, these Fe NCs are not textured. This
magnetic anisotropy could be due to the shape effect of Fe NCs,
\ie~they are not sphere-like, or magnetostriction. There is,
however, no evidence for any of these two possibilities.

\begin{figure} \center
\includegraphics[scale=0.7]{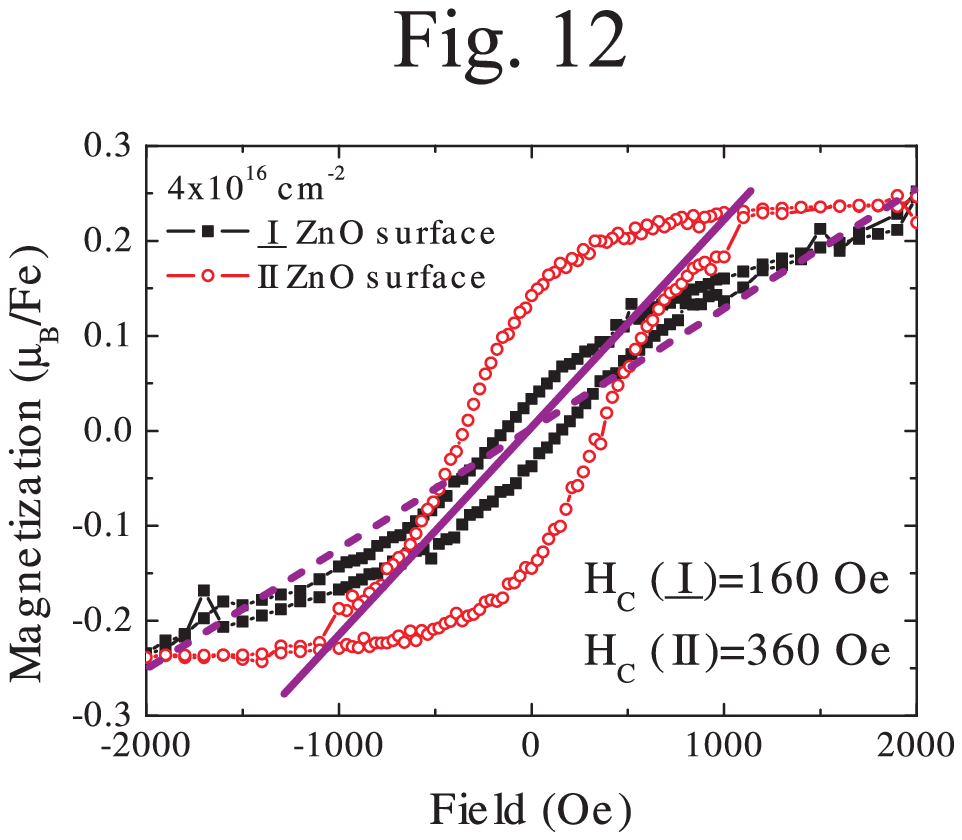}
\caption{Hysteresis loops measured at 5 K for Fe implanted ZnO at
180 keV and 623 K up to a fluence of 4$\times$10$^{16}$ cm$^{-2}$.
The field is changed from parallel to perpendicular with respect
to the sample surface, revealing the magnetic anisotropy. The
intersections between the easy axis M-H curve and solid and dashed
direct lines indicate the lowest and highest
$H_A$.}\label{fig:MH_FeZnO_aniso}
\end{figure}

\subsubsection{Memory effect of Fe NCs}

\begin{figure} \center
\includegraphics[scale=0.7]{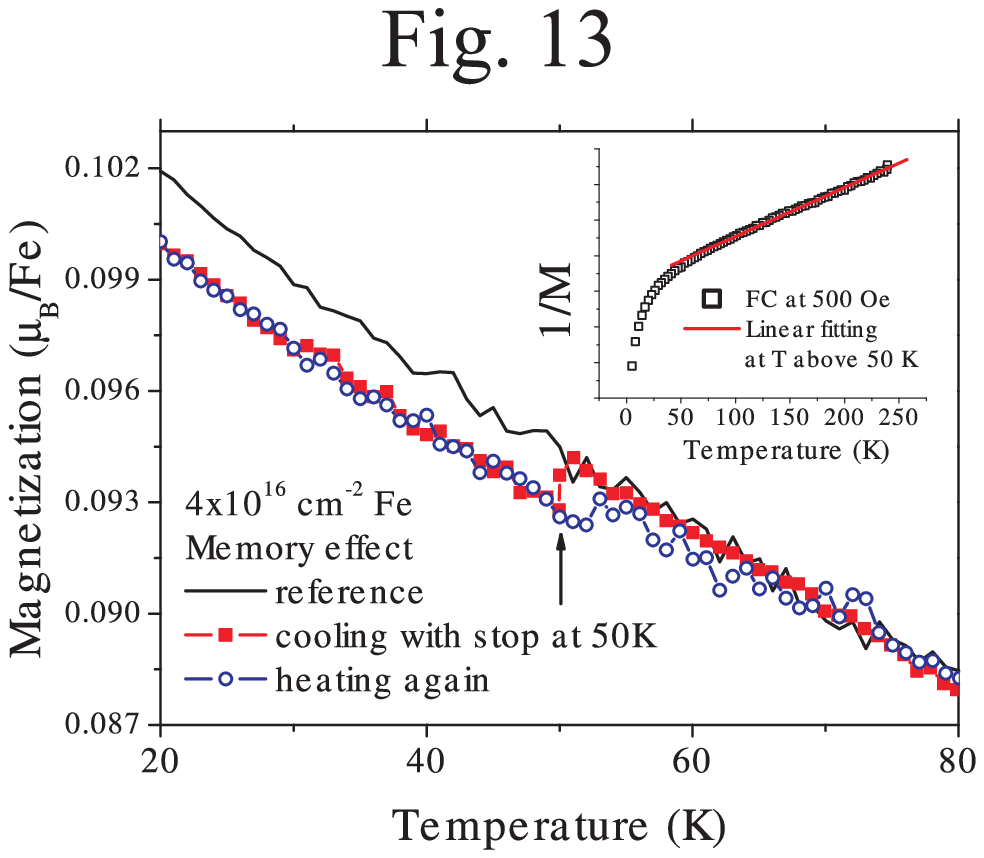}
\caption{Temperature dependent memory effect in the dc
magnetization. The reference curve is measured on heating at a
constant rage of 3 K/min after FC in 50 Oe. The solid squires are
measured during cooling in 50 Oe at the same rate but with  a stop
of 2 hours at 50 K. The field is cut off during stop. The open
circles are measured with continuous heating at the same rate
after the previous cooling protocol. Inset shows the reciprocal
magnetization versus temperature.}\label{fig:memory_effect}
\end{figure}

Below the blocking temperature, a magnetic nanoparticle system has
a rich and unusual behavior. For instance a slow relaxation and a
history-dependent magnetic memory are found in the dc
magnetization as a function of temperature
\cite{PhysRevLett.91.167206}\cite{tsoi:014445}\cite{zheng:139702}\cite{sasaki:104405}\cite{chakraverty:042501}.
In our system, Fe nanoparticles embedded inside ZnO crystals, the
temperature dependent memory effect was also observed (Figure 13)
using a cooling and heating protocol suggested by Sun
\etal~\cite{PhysRevLett.91.167206}. At 300 K a magnetic field of
50 Oe was applied and the sample was cooled down to 5 K at a
constant cooling rate of 3 K/min. Then the sample was heated
continuously at the same rate and the magnetization was recorded.
The obtained M(T) curve is referred as the reference curve (solid
line in Figure 13). Thereafter, we cooled the sample at the same
rate and recorded the magnetization with cooling, but temporarily
stopped at T = 50 K for a waiting time of 2 hours. During waiting
time, the field was switched off. After the stop, the 50 Oe field
was reapplied and cooling and measuring were resumed. The
temporary stop resulted in a steplike M(T) curve (solid squares in
Figure 13). After reaching the lowest temperature 5 K, the sample
was heated back with the rate of 3K/min in the same field, and the
magnetization was recorded again. The M(T) curve during this
heating also has a steplike behavior at the stop temperature of 50
K, then recovers the previous M(T) curve measured during cooling.
The system remembers its thermal history. Two explanations have
been suggested for such a memory effect \cite{sasaki:104405}. The
first one is a broad distribution of blocking temperatures
originating from the distribution of the anisotropy energy
barriers. Another explanation is the strong dipolar interaction
between nanoparticles, which frustrates the nanomagnetic moments,
and slows down their relaxation. Our observations rather support
the first model. First of all, the memory effect is also observed
for two other samples (2$\times$10$^{16}$, and 8$\times$10$^{16}$
cm$^{-2}$) (not shown). Therefore the effect is independent of ion
fluence, \ie~particle density. Second, the inset of Figure 13
shows the reciprocal FC magnetization at 500 Oe versus
temperature. The perfect linearity of the curve for T$>$50 K
strongly suggests that the dynamics of the nanoparticles above
blocking temperature can be well described by superparamagnetism.
Therefore, the magnetic properties of the sample depend only on
the individual particle behavior. Third, the size of Fe
nanoparticles is widely dispersed according to the analysis on the
ZFC magnetization curve as shown in Figure 8. Therefore, we would
attribute the memory effect to the broad distribution of particle
size, \ie~of anisotropy energy barriers.

\subsection{Fe implanted epitaxial ZnO layers}
The epitaxial ZnO layers used in this study were grown by pulsed
layer deposition on Al$_2$O$_3$(0001). These thin films are n-type
conducting with a carrier concentration of 10$^{15}$-10$^{17}$
cm$^3$ at room temperature. Details about the sample preparation
can be found in Refs.
\cite{lorenz}\cite{lorenz03}\cite{lorenz:243510}. $^{57}$Fe ions
were implanted at an energy of 180 keV at 623 K. Then the samples
were subjected to the same thermal annealing, and structural as
well as magnetic characterization like the bulk crystals.

\subsubsection{Formation of Fe NCs}

\begin{figure} \center
\includegraphics[scale=0.70]{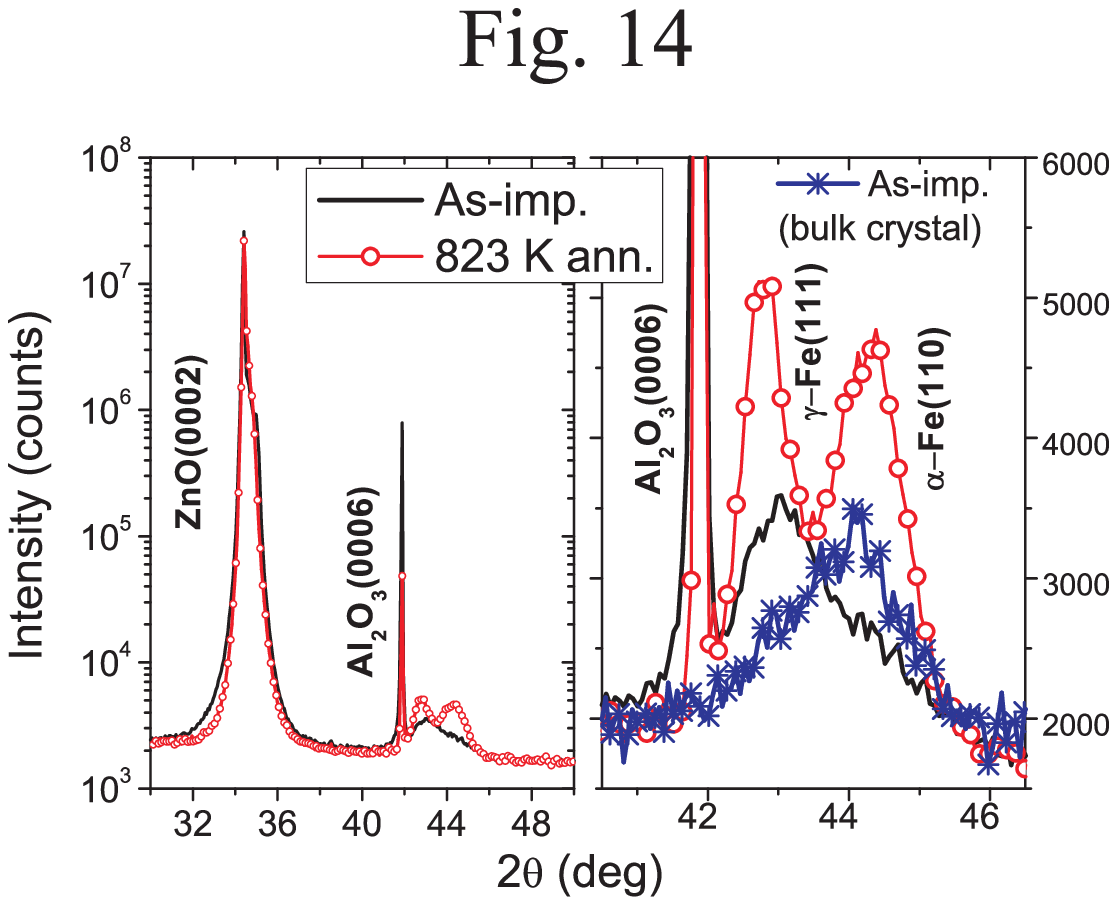}
\caption{XRD 2$\theta$-$\theta$ scans of Fe implanted epitaxial
ZnO layers: as-implanted and post annealed at 823 K. The bulk
crystal ZnO implanted at the same condition is shown for
comparison.}\label{fig:XRD_FeZnO_epi}
\end{figure}

Figure 14 shows the XRD 2$\theta$-$\theta$
scans of Fe implanted ZnO epitaxial layers (as-implanted and post
annealed at 823 K). In the left panel, one can see the nice
epitaxy of ZnO on Al$_2$O$_3$ with the out-of-plane relationship
of ZnO(0001)$\parallel$Al$_2$O$_3$(0001). The right panel is a
zoom on the Fe-related peak region. A single crystal sample
implanted with the same fluence and at the same temperature is
shown for comparison. Obviously, the epitaxial ZnO behaves
differently from the bulk crystals upon Fe implantation. In the
epitaxial-layer, $\gamma$-Fe is the predominant phase, while it is
$\alpha$-Fe in the single crystal. This difference will be
discussed in section \ref{section:discussions}. Upon thermal
annealing at 823 K, the epitaxial ZnO behaves similar to the low
temperature implanted bulk crystals. Both metallic Fe phases
($\alpha$ and $\gamma$) are growing.

\subsubsection{Charge state of Fe}

\begin{figure} \center
\includegraphics[scale=0.6]{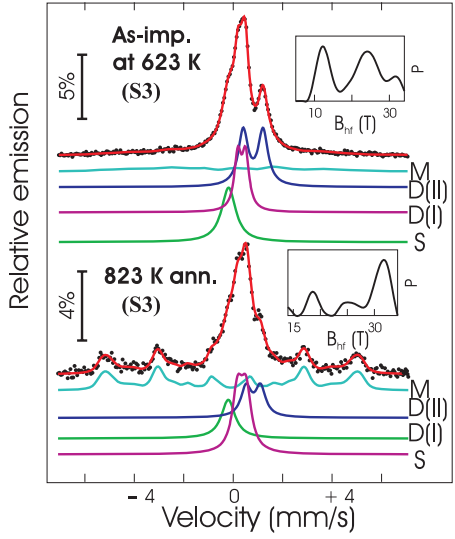}
\caption{Room temperature CEMS for ZnO epitaxial thin films
as-implanted with $^{57}$Fe and post-annealed at 823 K for 15 min.
The notations for the fitting lines are given as S (singlet), D
(doublet) and M (sextet). On the right side of the spectra, the
probability distribution P for the magnetic hyperfine field
(B$_{hf}$, solid lines) are given.}\label{fig:CEMS_epi}
\end{figure}

The charge and chemical states of Fe deduced from CEMS are shown
in Figure 15. The hyperfine interaction parameters are given in
Table \ref{tab:hyperfine_FeZnO}. In the as-implanted sample, ionic
Fe is the predominant phase, while also $\alpha$- and $\gamma$-Fe
are present (sextet M and singlet S, respectively). After
annealing at 823 K, the fraction of ferromagnetic $\alpha$-Fe is
drastically increased from 13.4\% to 38.8\%. In the bulk crystal
implanted at the same condition, there is no $\gamma$-Fe neither
in as-implanted nor in annealed samples. Also the fraction of
$\alpha$-Fe after annealing (18.2\%) is much lower than that in
epitaxial films (38.8\%).

\subsubsection{Magnetic properties}

\begin{figure} \center
\includegraphics[scale=0.70]{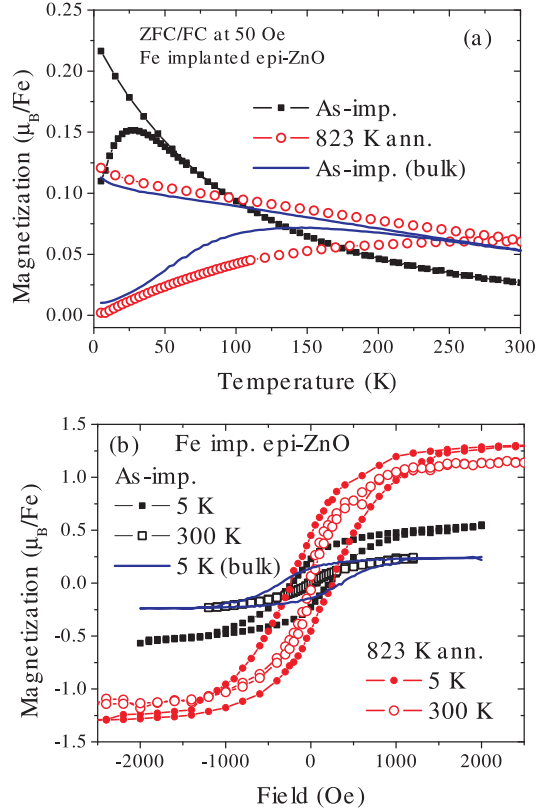}
\caption{(a) ZFC/FC magnetization curves at an applied field of 50
Oe for the as-implanted and 823 K annealed ZnO films; (b) M-H
loops at 5 K and 300 K. The bulk crystal ZnO implanted at the same
condition is shown as solid lines for
comparison.}\label{fig:SQUID_FeZnO_epi}
\end{figure}

The ZFC/FC magnetization measurements for the as-implanted and 823
K annealed samples (Figure 16(a)) shows the typical behavior of a
magnetic nanoparticle system. However T$_{B}$ in the ZFC curves
increases from 26 K to around 300 K with post annealing. The
absolute magnetization value per Fe in the ZFC/FC curves for the
annealed sample is lower than that of the as-implanted sample.
This is due to the fact that there are more bigger Fe NCs after
annealing, and the bigger NCs are more difficult to be aligned at
such a small field of 50 Oe. Figure 16(b) shows the M-H curves.
The coercivity is not significantly changed with annealing, while
the saturation magnetization is increased from 0.55$\mu_B$/Fe to
1.3$\mu_B$/Fe at 5 K and from 0.24$\mu_B$/Fe to 1.1$\mu_B$/Fe at
300 K, respectively, with annealing. For both samples, the M-H
loops show no hysteresis at 300 K without coercivity and
remanence. Obviously, the annealing behavior is different from the
single crystal implanted at same temperature of 623 K, but similar
to the single crystal implanted at 253 K. We will discuss this
point in section \ref{section:discussions}.

\section{Discussions} \label{section:discussions}

\subsection{Phase diagram of Fe in ZnO}
In section \ref{section:results}, we present the structure and
magnetic properties of Fe implanted ZnO. The implantation
parameters, \ie~fluence, energy, temperature, were varied. In
general, metallic Fe NCs have been formed already in the
as-implanted state when the implantation temperature is above 623
K and the fluence is above $2\times10^{16}$ cm$^{-2}$. By
summarizing all results, a phase diagram of Fe in ZnO can be
sketched, as shown in Figure 17. Note that the materials studied
in this research are ZnO bulk crystals grown by hydro-thermal
method. They are semi-insulating in the as-purchased state with
n-type carrier concentration of 10$^{12}$-10$^{14}$ cm$^{-3}$. The
phase diagram will likely be different for epitaxial-ZnO and for
p-type ZnO.

\begin{figure} \center
\includegraphics[scale=0.8]{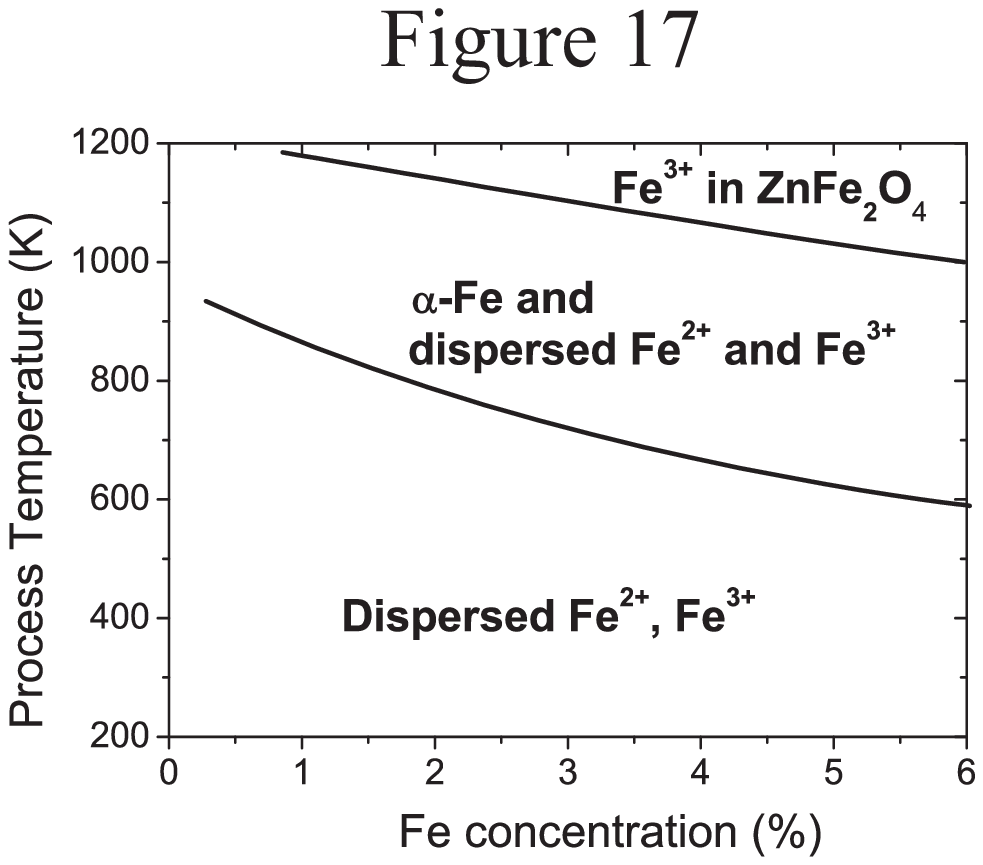}
\caption{The phase diagram of Fe in ZnO bulk crystals derived from
the data presented in this work. The process temperature refers to
the implantation or annealing
temperature.}\label{fig:phase_diagram}
\end{figure}

\subsection{Phase separation depends on the forms of ZnO}
In section \ref{section:results}, we have shown the structural and
magnetic properties of ZnO bulk crystals and epitaxial thin films
implanted at the same temperature and with the same Fe fluence.
They are obviously different from each other (see table
\ref{tab:FeZnO}). In the bulk crystals, only 11\% (increased to
15\% after 823 K post annealing) of the implanted Fe is
ferromagnetic and mainly $\alpha$-Fe, while 25\% in the epitaxial
ZnO films (increased to 59\% after 823 K post annealing). Recently
Dietl proposed the self-organized growth driven by the charge
state of the magnetic impurities \cite{dietlmat}\cite{dietlnatm}.
The energy levels derived from the open $d$ shells of transition
metals reside usually in the bandgap of the host semiconductor.
The mid-gap levels of magnetic impurities trap carriers
origination from residual impurities or defects. This trapping
alters the charge state of the magnetic ions and hence affects
their mutual Coulomb interactions. Therefore, different carriers
(electrons or holes, with different concentrations) could lead to
different interactions (\eg~repulsions and attractions) between
the implanted transition metal ions, and finally result in a
different phase separation. Both ZnO materials (bulk crystals and
epitaxial thin films) used in this study are n-type
semiconductors. The carrier concentration is around
10$^{12}$-10$^{13}$ cm$^{-3}$ for bulk crystals and
10$^{15}$-10$^{17}$ cm$^3$ for epitaxial thin films
\cite{lorenz}\cite{lorenz03} at room
temperature. 
Therefore, we can explain the different behavior in ZnO bulk
crystals and epitaxial layers upon Fe implantation in the
above-mentioned model. A higher concentration of free electrons
leads to more agglomerations of Fe. Moreover the nanocrystal
aggregation could be largely reduced or avoided by the realization
of p-type doping in ZnO.

However, one has to note that there are a lot of defects, such as
dislocations, and stacking faults, in the epitaxial ZnO films
grown on Al$_2$O$_3$ due to the large lattice mismatch
\cite{narayan:2597}. Kaiser \etal~demonstrated that in high
fluence Er implanted SiC, the defects act as nucleation sites in
the formation of Er-atom cluster and NCs \cite{Kaiser02}. A
similar effect can be present in the case of Fe implanted ZnO
films.

\subsection{Annealing behavior depends on the initial state}

Note that three kind of samples have been annealed at the same
temperature of 823 K. One is the ZnO single crystal implanted at
623 K, in which Fe NCs have already been formed in the
as-implanted states. One is the ZnO single crystal implanted at
253 K, in which no Fe NCs could be detected in the as-implanted
sample. The last one is the ZnO thin film implanted at 623 K, in
which rather small Fe NCs have been formed in the as-implanted
sample compared to that in the single crystal. The annealing
behavior in the first case is quite different from the later two
cases. One reason is the fact that diffusion of Fe NCs is much
more difficult than of single Fe ions. In the 623 K implanted
sample, Fe NCs have already been formed, and they are not so
mobile during 823 K annealing. Therefore the size and the amount
of Fe NCs only slight increase after annealing. However, in the
sample without Fe NCs or with very small NCs, Fe ions are more
mobile with annealing, and they aggregate into rather larger Fe
NCs. Another reason could be the same as discussed in the above
section, given the fact that implantation at 253 K induces more
point defects than that 623 K (see Figure 2).

\subsection{Magnetic coupling of dispersed ionic Fe}

Although a part of the implanted Fe ions aggregated to metallic
NCs, the remaining are in the ionic state. Even after 823 K
annealing, there is still a considerable amount of ionic Fe. By
SR-XRD no crystalline Fe-oxides could be detected. Therefore, these
ionic Fe could be diluted inside ZnO matrix. Moreover, Ref.
\cite{kucheyev02} demonstrated that the implantation-induced
electrical isolation of ZnO is removed after annealing between 773
to 873 K. Therefore, the carrier concentration is comparable with
the virgin sample after annealing at 823 K. However, as measured by
SQUID down to 5 K and CEMS at room temperature, Fe$^{2+}$ and
Fe$^{3+}$ are not ferromagnetically coupled.

In addition to conventional thermal annealing which is an
equilibrium process, a nonequilibrium annealing technique, \ie~flash lamp annealing at a pulse length of 20 ms, was also used by
us \cite{potzger07}. For an intermediate light power, the
implantation-induced surface defects could be removed without
creation of secondary phases within the implanted region. However,
there is still no detectable ferromagnetic coupling between these
dispersed Fe ions.

Moreover, currently the absence of ferromagnetism in transition
metal doped ZnO is a universal problem. Several groups have shown
that transition metal ions, \eg~Fe \cite{kolesnik:2582}, Mn
\cite{rao05}\cite{kolesnik:2582}, and Co
\cite{rao05}\cite{zhang06}\cite{kolesnik:2582}, are substitutional
inside ZnO. However, no ferromagnetism could be observed due to
the possible reason of the lack of p-type conductivity.

\section{Summary and Conclusions}\label{section:conclusions}

(1) In general, a combination of SR-XRD, ZFC/FC magnetization and
element specific spectroscopy measurements is a reliable approach
to clarify the observed magnetism in DMS materials.

(2) By correlating the structural and magnetic properties of all
investigated samples, it is clear that ferromagnetism is only
observed when $\alpha$-Fe (or ZnFe$_2$O$_4$) NCs are present. In
as-implanted and 823 K annealed samples, dispersed Fe$^{2+}$ and
Fe$^{3+}$ are the predominant charge states. However, they are not
ferromagnetically coupled.

(3) $\alpha$-Fe (bcc) NCs are not crystallographically oriented
inside ZnO matrix. However, fcc-ZnFe$_2$O$_4$ NCs formed after
annealing at 1073 K are epitaxially embedded in ZnO. This is due
to the crystalline symmetry. Hexagonal ZnO crystals are six-fold
symmetric, while $\alpha$-Fe is four-fold symmetric.
Fcc-ZnFe$_2$O$_4$ is also six-fold symmetric viewed along [111]
direction.

(4) The magnetic properties of these Fe NCs were carefully
investigated regarding their memory effect and magnetic
anisotropy. A memory effect is observed in the temperature
dependent magnetization measurement, which is induced by the
different relaxation times originating from the different grain
sizes of the Fe nanoparticles, and consequently different
anisotropy energy barriers. The in-plane magnetic anisotropy could
be due to the shape effect.

(5) The phase separation, \ie~the formation of metallic Fe,
depends on the initial state of the host materials, namely the
carrier and/or the  defect concentrations. The n-type carriers
could facilitate the self-organization of metallic Fe NCs.

(6) The next question is directed to the magnetically activation
of the diluted ionic Fe in ZnO. The realization of p-type doping
for ZnO could be the solution.

\section{Acknowledgments}
We are thankful Holger Hochmuth for the growth of epitaxial ZnO
films and Heidemarie Schmidt for fruitful discussions.


\begin{thebibliography}{76}
\expandafter\ifx\csname natexlab\endcsname\relax\def\natexlab#1{#1}\fi
\expandafter\ifx\csname bibnamefont\endcsname\relax
  \def\bibnamefont#1{#1}\fi
\expandafter\ifx\csname bibfnamefont\endcsname\relax
  \def\bibfnamefont#1{#1}\fi
\expandafter\ifx\csname citenamefont\endcsname\relax
  \def\citenamefont#1{#1}\fi
\expandafter\ifx\csname url\endcsname\relax
  \def\url#1{\texttt{#1}}\fi
\expandafter\ifx\csname urlprefix\endcsname\relax\def\urlprefix{URL }\fi
\providecommand{\bibinfo}[2]{#2}
\providecommand{\eprint}[2][]{\url{#2}}

\bibitem[{\citenamefont{Dietl et~al.}(2000)\citenamefont{Dietl, Ohno,
  Matsukura, Cibert, and Ferrand}}]{dietl00}
\bibinfo{author}{\bibfnamefont{T.}~\bibnamefont{Dietl}},
  \bibinfo{author}{\bibfnamefont{H.}~\bibnamefont{Ohno}},
  \bibinfo{author}{\bibfnamefont{F.}~\bibnamefont{Matsukura}},
  \bibinfo{author}{\bibfnamefont{J.}~\bibnamefont{Cibert}}, \bibnamefont{and}
  \bibinfo{author}{\bibfnamefont{D.}~\bibnamefont{Ferrand}},
  \bibinfo{journal}{Science} \textbf{\bibinfo{volume}{287}},
  \bibinfo{pages}{1019} (\bibinfo{year}{2000}).

\bibitem[{\citenamefont{Sato and Katayama-Yoshida}(2001)}]{sato_ZnO}
\bibinfo{author}{\bibfnamefont{K.}~\bibnamefont{Sato}} \bibnamefont{and}
  \bibinfo{author}{\bibfnamefont{H.}~\bibnamefont{Katayama-Yoshida}},
  \bibinfo{journal}{Physica E} \textbf{\bibinfo{volume}{10}},
  \bibinfo{pages}{251} (\bibinfo{year}{2001}).

\bibitem[{\citenamefont{Angadi et~al.}(2006)\citenamefont{Angadi, Jung, Choi,
  Kumar, Jeong, Shin, Lee, Song, Khan, and Srivastava}}]{angadi06}
\bibinfo{author}{\bibfnamefont{B.}~\bibnamefont{Angadi}},
  \bibinfo{author}{\bibfnamefont{Y.~S.} \bibnamefont{Jung}},
  \bibinfo{author}{\bibfnamefont{W.~K.} \bibnamefont{Choi}},
  \bibinfo{author}{\bibfnamefont{R.}~\bibnamefont{Kumar}},
  \bibinfo{author}{\bibfnamefont{K.}~\bibnamefont{Jeong}},
  \bibinfo{author}{\bibfnamefont{S.~W.} \bibnamefont{Shin}},
  \bibinfo{author}{\bibfnamefont{J.~H.} \bibnamefont{Lee}},
  \bibinfo{author}{\bibfnamefont{J.~H.} \bibnamefont{Song}},
  \bibinfo{author}{\bibfnamefont{M.~W.} \bibnamefont{Khan}}, \bibnamefont{and}
  \bibinfo{author}{\bibfnamefont{J.~P.} \bibnamefont{Srivastava}},
  \bibinfo{journal}{Appl. Phys. Lett.} \textbf{\bibinfo{volume}{88}},
  \bibinfo{pages}{142502} (\bibinfo{year}{2006}).

\bibitem[{\citenamefont{Heo et~al.}(2004)\citenamefont{Heo, Ivill, Ip, Norton,
  Pearton, Kelly, Rairigh, Hebard, and Steiner}}]{heo04}
\bibinfo{author}{\bibfnamefont{Y.~W.} \bibnamefont{Heo}},
  \bibinfo{author}{\bibfnamefont{M.~P.} \bibnamefont{Ivill}},
  \bibinfo{author}{\bibfnamefont{K.}~\bibnamefont{Ip}},
  \bibinfo{author}{\bibfnamefont{D.~P.} \bibnamefont{Norton}},
  \bibinfo{author}{\bibfnamefont{S.~J.} \bibnamefont{Pearton}},
  \bibinfo{author}{\bibfnamefont{J.~G.} \bibnamefont{Kelly}},
  \bibinfo{author}{\bibfnamefont{R.}~\bibnamefont{Rairigh}},
  \bibinfo{author}{\bibfnamefont{A.~F.} \bibnamefont{Hebard}},
  \bibnamefont{and} \bibinfo{author}{\bibfnamefont{T.}~\bibnamefont{Steiner}},
  \bibinfo{journal}{Appl. Phys. Lett.} \textbf{\bibinfo{volume}{84}},
  \bibinfo{pages}{2292} (\bibinfo{year}{2004}).

\bibitem[{\citenamefont{Hong et~al.}(2005{\natexlab{a}})\citenamefont{Hong,
  Brize, and Sakai}}]{hong05}
\bibinfo{author}{\bibfnamefont{N.~H.} \bibnamefont{Hong}},
  \bibinfo{author}{\bibfnamefont{V.}~\bibnamefont{Brize}}, \bibnamefont{and}
  \bibinfo{author}{\bibfnamefont{J.}~\bibnamefont{Sakai}},
  \bibinfo{journal}{Appl. Phys. Lett.} \textbf{\bibinfo{volume}{86}},
  \bibinfo{pages}{082505} (\bibinfo{year}{2005}{\natexlab{a}}).

\bibitem[{\citenamefont{Hong et~al.}(2005{\natexlab{b}})\citenamefont{Hong,
  Sakai, and Hassini}}]{hongv}
\bibinfo{author}{\bibfnamefont{N.~H.} \bibnamefont{Hong}},
  \bibinfo{author}{\bibfnamefont{J.}~\bibnamefont{Sakai}}, \bibnamefont{and}
  \bibinfo{author}{\bibfnamefont{A.}~\bibnamefont{Hassini}},
  \bibinfo{journal}{J. Phys.-Condens. Matter} \textbf{\bibinfo{volume}{17}},
  \bibinfo{pages}{199} (\bibinfo{year}{2005}{\natexlab{b}}).

\bibitem[{\citenamefont{Ip et~al.}(2003)\citenamefont{Ip, Frazier, Heo, Norton,
  Abernathy, Pearton, Kelly, Rairigh, Hebard, Zavada et~al.}}]{ip03}
\bibinfo{author}{\bibfnamefont{K.}~\bibnamefont{Ip}},
  \bibinfo{author}{\bibfnamefont{R.~M.} \bibnamefont{Frazier}},
  \bibinfo{author}{\bibfnamefont{Y.~W.} \bibnamefont{Heo}},
  \bibinfo{author}{\bibfnamefont{D.~P.} \bibnamefont{Norton}},
  \bibinfo{author}{\bibfnamefont{C.~R.} \bibnamefont{Abernathy}},
  \bibinfo{author}{\bibfnamefont{S.~J.} \bibnamefont{Pearton}},
  \bibinfo{author}{\bibfnamefont{J.}~\bibnamefont{Kelly}},
  \bibinfo{author}{\bibfnamefont{R.}~\bibnamefont{Rairigh}},
  \bibinfo{author}{\bibfnamefont{A.~F.} \bibnamefont{Hebard}},
  \bibinfo{author}{\bibfnamefont{J.~M.} \bibnamefont{Zavada}},
  \bibnamefont{et~al.}, \bibinfo{journal}{J. Vac. Sci. Technol. B}
  \textbf{\bibinfo{volume}{21}}, \bibinfo{pages}{1476} (\bibinfo{year}{2003}).

\bibitem[{\citenamefont{Jung et~al.}(2002)\citenamefont{Jung, An, Yi, Jung,
  Lee, and Cho}}]{jung02}
\bibinfo{author}{\bibfnamefont{S.~W.} \bibnamefont{Jung}},
  \bibinfo{author}{\bibfnamefont{S.~J.} \bibnamefont{An}},
  \bibinfo{author}{\bibfnamefont{G.~C.} \bibnamefont{Yi}},
  \bibinfo{author}{\bibfnamefont{C.~U.} \bibnamefont{Jung}},
  \bibinfo{author}{\bibfnamefont{S.~I.} \bibnamefont{Lee}}, \bibnamefont{and}
  \bibinfo{author}{\bibfnamefont{S.}~\bibnamefont{Cho}},
  \bibinfo{journal}{Appl. Phys. Lett.} \textbf{\bibinfo{volume}{80}},
  \bibinfo{pages}{4561} (\bibinfo{year}{2002}).

\bibitem[{\citenamefont{Polyakov et~al.}(2004)\citenamefont{Polyakov, Govorkov,
  Smirnov, Pashkova, Pearton, Ip, Frazier, Abernathy, Norton, Zavada
  et~al.}}]{pol04}
\bibinfo{author}{\bibfnamefont{A.~Y.} \bibnamefont{Polyakov}},
  \bibinfo{author}{\bibfnamefont{A.~V.} \bibnamefont{Govorkov}},
  \bibinfo{author}{\bibfnamefont{N.~B.} \bibnamefont{Smirnov}},
  \bibinfo{author}{\bibfnamefont{N.~V.} \bibnamefont{Pashkova}},
  \bibinfo{author}{\bibfnamefont{S.~J.} \bibnamefont{Pearton}},
  \bibinfo{author}{\bibfnamefont{K.}~\bibnamefont{Ip}},
  \bibinfo{author}{\bibfnamefont{R.~M.} \bibnamefont{Frazier}},
  \bibinfo{author}{\bibfnamefont{C.~R.} \bibnamefont{Abernathy}},
  \bibinfo{author}{\bibfnamefont{D.~P.} \bibnamefont{Norton}},
  \bibinfo{author}{\bibfnamefont{J.~M.} \bibnamefont{Zavada}},
  \bibnamefont{et~al.}, \bibinfo{journal}{Mater. Sci. Semicond. Process}
  \textbf{\bibinfo{volume}{7}}, \bibinfo{pages}{77} (\bibinfo{year}{2004}).

\bibitem[{\citenamefont{Tuan et~al.}(2004)\citenamefont{Tuan, Bryan, Pakhomov,
  Shutthanandan, Thevuthasan, McCready, Gaspar, Engelhard, Rogers, Krishnan
  et~al.}}]{tuan04}
\bibinfo{author}{\bibfnamefont{A.~C.} \bibnamefont{Tuan}},
  \bibinfo{author}{\bibfnamefont{J.~D.} \bibnamefont{Bryan}},
  \bibinfo{author}{\bibfnamefont{A.~B.} \bibnamefont{Pakhomov}},
  \bibinfo{author}{\bibfnamefont{V.}~\bibnamefont{Shutthanandan}},
  \bibinfo{author}{\bibfnamefont{S.}~\bibnamefont{Thevuthasan}},
  \bibinfo{author}{\bibfnamefont{D.~E.} \bibnamefont{McCready}},
  \bibinfo{author}{\bibfnamefont{D.}~\bibnamefont{Gaspar}},
  \bibinfo{author}{\bibfnamefont{M.~H.} \bibnamefont{Engelhard}},
  \bibinfo{author}{\bibfnamefont{J.~W.} \bibnamefont{Rogers}},
  \bibinfo{author}{\bibfnamefont{K.}~\bibnamefont{Krishnan}},
  \bibnamefont{et~al.}, \bibinfo{journal}{Phys. Rev. B}
  \textbf{\bibinfo{volume}{70}}, \bibinfo{pages}{054424}
  (\bibinfo{year}{2004}).

\bibitem[{\citenamefont{Venkatesan et~al.}(2004)\citenamefont{Venkatesan,
  Fitzgerald, Lunney, and Coey}}]{venk04}
\bibinfo{author}{\bibfnamefont{M.}~\bibnamefont{Venkatesan}},
  \bibinfo{author}{\bibfnamefont{C.~B.} \bibnamefont{Fitzgerald}},
  \bibinfo{author}{\bibfnamefont{J.~G.} \bibnamefont{Lunney}},
  \bibnamefont{and} \bibinfo{author}{\bibfnamefont{J.~M.~D.}
  \bibnamefont{Coey}}, \bibinfo{journal}{Phys. Rev. Lett.}
  \textbf{\bibinfo{volume}{93}}, \bibinfo{pages}{177206}
  (\bibinfo{year}{2004}).

\bibitem[{\citenamefont{Bouloudenine et~al.}(2005)\citenamefont{Bouloudenine,
  Viart, Colis, Kortus, and Dinia}}]{boul05}
\bibinfo{author}{\bibfnamefont{M.}~\bibnamefont{Bouloudenine}},
  \bibinfo{author}{\bibfnamefont{N.}~\bibnamefont{Viart}},
  \bibinfo{author}{\bibfnamefont{S.}~\bibnamefont{Colis}},
  \bibinfo{author}{\bibfnamefont{J.}~\bibnamefont{Kortus}}, \bibnamefont{and}
  \bibinfo{author}{\bibfnamefont{A.}~\bibnamefont{Dinia}},
  \bibinfo{journal}{Appl. Phys. Lett.} \textbf{\bibinfo{volume}{87}},
  \bibinfo{pages}{052501} (\bibinfo{year}{2005}).

\bibitem[{\citenamefont{Yin et~al.}(2006)\citenamefont{Yin, Xu, Yang, Liu,
  Rosner, Hahn, Gleiter, Schild, Doyle, Liu et~al.}}]{yin06}
\bibinfo{author}{\bibfnamefont{S.}~\bibnamefont{Yin}},
  \bibinfo{author}{\bibfnamefont{M.~X.} \bibnamefont{Xu}},
  \bibinfo{author}{\bibfnamefont{L.}~\bibnamefont{Yang}},
  \bibinfo{author}{\bibfnamefont{J.~F.} \bibnamefont{Liu}},
  \bibinfo{author}{\bibfnamefont{H.}~\bibnamefont{Rosner}},
  \bibinfo{author}{\bibfnamefont{H.}~\bibnamefont{Hahn}},
  \bibinfo{author}{\bibfnamefont{H.}~\bibnamefont{Gleiter}},
  \bibinfo{author}{\bibfnamefont{D.}~\bibnamefont{Schild}},
  \bibinfo{author}{\bibfnamefont{S.}~\bibnamefont{Doyle}},
  \bibinfo{author}{\bibfnamefont{T.}~\bibnamefont{Liu}}, \bibnamefont{et~al.},
  \bibinfo{journal}{Phys. Rev. B} \textbf{\bibinfo{volume}{73}},
  \bibinfo{pages}{224408} (\bibinfo{year}{2006}).

\bibitem[{\citenamefont{Sati et~al.}(2007)\citenamefont{Sati, Deparis, Morhain,
  Schafer, and Stepanov}}]{sati:137204}
\bibinfo{author}{\bibfnamefont{P.}~\bibnamefont{Sati}},
  \bibinfo{author}{\bibfnamefont{C.}~\bibnamefont{Deparis}},
  \bibinfo{author}{\bibfnamefont{C.}~\bibnamefont{Morhain}},
  \bibinfo{author}{\bibfnamefont{S.}~\bibnamefont{Schafer}}, \bibnamefont{and}
  \bibinfo{author}{\bibfnamefont{A.}~\bibnamefont{Stepanov}},
  \bibinfo{journal}{Phys. Rev. Lett.} \textbf{\bibinfo{volume}{98}},
  \bibinfo{pages}{137204} (\bibinfo{year}{2007}).

\bibitem[{\citenamefont{Fukumura et~al.}(2001)\citenamefont{Fukumura, Jin, ,
  Kawasaki, Shono, Hasegawa, Koshihara, and Koinuma}}]{fukumura01}
\bibinfo{author}{\bibfnamefont{M.}~\bibnamefont{Fukumura}},
  \bibinfo{author}{\bibfnamefont{Z.~W.} \bibnamefont{Jin}}, ,
  \bibinfo{author}{\bibfnamefont{M.}~\bibnamefont{Kawasaki}},
  \bibinfo{author}{\bibfnamefont{T.}~\bibnamefont{Shono}},
  \bibinfo{author}{\bibfnamefont{T.}~\bibnamefont{Hasegawa}},
  \bibinfo{author}{\bibfnamefont{S.}~\bibnamefont{Koshihara}},
  \bibnamefont{and} \bibinfo{author}{\bibfnamefont{H.}~\bibnamefont{Koinuma}},
  \bibinfo{journal}{Appl. Phys. Lett.} \textbf{\bibinfo{volume}{78}},
  \bibinfo{pages}{958} (\bibinfo{year}{2001}).

\bibitem[{\citenamefont{Jin et~al.}(2001)\citenamefont{Jin, Fukumura, Kawasaki,
  Ando, Saito, Sekiguchi, Yoo, Murakami, Matsumoto, Hasegawa et~al.}}]{jin01}
\bibinfo{author}{\bibfnamefont{Z.~W.} \bibnamefont{Jin}},
  \bibinfo{author}{\bibfnamefont{T.}~\bibnamefont{Fukumura}},
  \bibinfo{author}{\bibfnamefont{M.}~\bibnamefont{Kawasaki}},
  \bibinfo{author}{\bibfnamefont{K.}~\bibnamefont{Ando}},
  \bibinfo{author}{\bibfnamefont{H.}~\bibnamefont{Saito}},
  \bibinfo{author}{\bibfnamefont{T.}~\bibnamefont{Sekiguchi}},
  \bibinfo{author}{\bibfnamefont{Y.~Z.} \bibnamefont{Yoo}},
  \bibinfo{author}{\bibfnamefont{M.}~\bibnamefont{Murakami}},
  \bibinfo{author}{\bibfnamefont{Y.}~\bibnamefont{Matsumoto}},
  \bibinfo{author}{\bibfnamefont{T.}~\bibnamefont{Hasegawa}},
  \bibnamefont{et~al.}, \bibinfo{journal}{Appl. Phys. Lett.}
  \textbf{\bibinfo{volume}{78}}, \bibinfo{pages}{3824} (\bibinfo{year}{2001}).

\bibitem[{\citenamefont{Rao and Deepak}(2005)}]{rao05}
\bibinfo{author}{\bibfnamefont{C.~N.~R.} \bibnamefont{Rao}} \bibnamefont{and}
  \bibinfo{author}{\bibfnamefont{F.~L.} \bibnamefont{Deepak}},
  \bibinfo{journal}{J. Mater. Chem.} \textbf{\bibinfo{volume}{15}},
  \bibinfo{pages}{573} (\bibinfo{year}{2005}).

\bibitem[{\citenamefont{Zhang et~al.}(2006)\citenamefont{Zhang, Chen, Lee, Xue,
  Sun, Chen, Chen, and Chu}}]{zhang06}
\bibinfo{author}{\bibfnamefont{Z.}~\bibnamefont{Zhang}},
  \bibinfo{author}{\bibfnamefont{Q.}~\bibnamefont{Chen}},
  \bibinfo{author}{\bibfnamefont{H.~D.} \bibnamefont{Lee}},
  \bibinfo{author}{\bibfnamefont{Y.~Y.} \bibnamefont{Xue}},
  \bibinfo{author}{\bibfnamefont{Y.~Y.} \bibnamefont{Sun}},
  \bibinfo{author}{\bibfnamefont{H.}~\bibnamefont{Chen}},
  \bibinfo{author}{\bibfnamefont{F.}~\bibnamefont{Chen}}, \bibnamefont{and}
  \bibinfo{author}{\bibfnamefont{W.~K.} \bibnamefont{Chu}},
  \bibinfo{journal}{J. Appl. Phys.} \textbf{\bibinfo{volume}{100}},
  \bibinfo{pages}{043909} (\bibinfo{year}{2006}).

\bibitem[{\citenamefont{Norton et~al.}(2003)\citenamefont{Norton, Overberg,
  Pearton, Pruessner, Budai, Boatner, Chisholm, Lee, Khim, Park
  et~al.}}]{norton03}
\bibinfo{author}{\bibfnamefont{D.~P.} \bibnamefont{Norton}},
  \bibinfo{author}{\bibfnamefont{M.~E.} \bibnamefont{Overberg}},
  \bibinfo{author}{\bibfnamefont{S.~J.} \bibnamefont{Pearton}},
  \bibinfo{author}{\bibfnamefont{K.}~\bibnamefont{Pruessner}},
  \bibinfo{author}{\bibfnamefont{J.~D.} \bibnamefont{Budai}},
  \bibinfo{author}{\bibfnamefont{L.~A.} \bibnamefont{Boatner}},
  \bibinfo{author}{\bibfnamefont{M.~F.} \bibnamefont{Chisholm}},
  \bibinfo{author}{\bibfnamefont{J.~S.} \bibnamefont{Lee}},
  \bibinfo{author}{\bibfnamefont{Z.~G.} \bibnamefont{Khim}},
  \bibinfo{author}{\bibfnamefont{Y.~D.} \bibnamefont{Park}},
  \bibnamefont{et~al.}, \bibinfo{journal}{Appl. Phys. Lett.}
  \textbf{\bibinfo{volume}{83}}, \bibinfo{pages}{5488} (\bibinfo{year}{2003}).

\bibitem[{\citenamefont{Park et~al.}(2004)\citenamefont{Park, Kim, Jang, Ryu,
  and Kim}}]{park04}
\bibinfo{author}{\bibfnamefont{J.~H.} \bibnamefont{Park}},
  \bibinfo{author}{\bibfnamefont{M.~G.} \bibnamefont{Kim}},
  \bibinfo{author}{\bibfnamefont{H.~M.} \bibnamefont{Jang}},
  \bibinfo{author}{\bibfnamefont{S.}~\bibnamefont{Ryu}}, \bibnamefont{and}
  \bibinfo{author}{\bibfnamefont{Y.~M.} \bibnamefont{Kim}},
  \bibinfo{journal}{Appl. Phys. Lett.} \textbf{\bibinfo{volume}{84}},
  \bibinfo{pages}{1338} (\bibinfo{year}{2004}).

\bibitem[{\citenamefont{Kundaliya et~al.}(2004)\citenamefont{Kundaliya, Ogale,
  Lofland, Dhar, Metting, Shinde, Ma, Varughese, Ramanujachary, Salamanca-Riba
  et~al.}}]{kund04}
\bibinfo{author}{\bibfnamefont{D.~C.} \bibnamefont{Kundaliya}},
  \bibinfo{author}{\bibfnamefont{S.~B.} \bibnamefont{Ogale}},
  \bibinfo{author}{\bibfnamefont{S.~E.} \bibnamefont{Lofland}},
  \bibinfo{author}{\bibfnamefont{S.}~\bibnamefont{Dhar}},
  \bibinfo{author}{\bibfnamefont{C.~J.} \bibnamefont{Metting}},
  \bibinfo{author}{\bibfnamefont{S.~R.} \bibnamefont{Shinde}},
  \bibinfo{author}{\bibfnamefont{Z.}~\bibnamefont{Ma}},
  \bibinfo{author}{\bibfnamefont{B.}~\bibnamefont{Varughese}},
  \bibinfo{author}{\bibfnamefont{K.~V.} \bibnamefont{Ramanujachary}},
  \bibinfo{author}{\bibfnamefont{L.}~\bibnamefont{Salamanca-Riba}},
  \bibnamefont{et~al.}, \bibinfo{journal}{Nat. Mater.}
  \textbf{\bibinfo{volume}{3}}, \bibinfo{pages}{709} (\bibinfo{year}{2004}).

\bibitem[{\citenamefont{Shim et~al.}(2005)\citenamefont{Shim, Hwang, Lee, Park,
  Han, and Jeong}}]{shim05}
\bibinfo{author}{\bibfnamefont{J.~H.} \bibnamefont{Shim}},
  \bibinfo{author}{\bibfnamefont{T.}~\bibnamefont{Hwang}},
  \bibinfo{author}{\bibfnamefont{S.}~\bibnamefont{Lee}},
  \bibinfo{author}{\bibfnamefont{J.~H.} \bibnamefont{Park}},
  \bibinfo{author}{\bibfnamefont{S.~J.} \bibnamefont{Han}}, \bibnamefont{and}
  \bibinfo{author}{\bibfnamefont{Y.~H.} \bibnamefont{Jeong}},
  \bibinfo{journal}{Appl. Phys. Lett.} \textbf{\bibinfo{volume}{86}},
  \bibinfo{pages}{082503} (\bibinfo{year}{2005}).

\bibitem[{\citenamefont{Potzger et~al.}(2006)\citenamefont{Potzger, Zhou,
  Reuther, M\"{u}cklich, Eichhorn, Schell, Skorupa, Helm, Fassbender,
  Herrmannsdorfer et~al.}}]{pot06fe}
\bibinfo{author}{\bibfnamefont{K.}~\bibnamefont{Potzger}},
  \bibinfo{author}{\bibfnamefont{S.~Q.} \bibnamefont{Zhou}},
  \bibinfo{author}{\bibfnamefont{H.}~\bibnamefont{Reuther}},
  \bibinfo{author}{\bibfnamefont{A.}~\bibnamefont{M\"{u}cklich}},
  \bibinfo{author}{\bibfnamefont{F.}~\bibnamefont{Eichhorn}},
  \bibinfo{author}{\bibfnamefont{N.}~\bibnamefont{Schell}},
  \bibinfo{author}{\bibfnamefont{W.}~\bibnamefont{Skorupa}},
  \bibinfo{author}{\bibfnamefont{M.}~\bibnamefont{Helm}},
  \bibinfo{author}{\bibfnamefont{J.}~\bibnamefont{Fassbender}},
  \bibinfo{author}{\bibfnamefont{T.}~\bibnamefont{Herrmannsdorfer}},
  \bibnamefont{et~al.}, \bibinfo{journal}{Appl. Phys. Lett.}
  \textbf{\bibinfo{volume}{88}}, \bibinfo{pages}{052508}
  (\bibinfo{year}{2006}).

\bibitem[{\citenamefont{Shinagawa et~al.}(2006)\citenamefont{Shinagawa, Izaki,
  Inui, Murase, and Awakura}}]{shin06}
\bibinfo{author}{\bibfnamefont{T.}~\bibnamefont{Shinagawa}},
  \bibinfo{author}{\bibfnamefont{M.}~\bibnamefont{Izaki}},
  \bibinfo{author}{\bibfnamefont{H.}~\bibnamefont{Inui}},
  \bibinfo{author}{\bibfnamefont{K.}~\bibnamefont{Murase}}, \bibnamefont{and}
  \bibinfo{author}{\bibfnamefont{Y.}~\bibnamefont{Awakura}},
  \bibinfo{journal}{Chem. Mater.} \textbf{\bibinfo{volume}{18}},
  \bibinfo{pages}{763} (\bibinfo{year}{2006}).

\bibitem[{\citenamefont{Talut et~al.}(2006)\citenamefont{Talut, Reuther,
  M\"{u}cklich, Eichhorn, and Potzger}}]{talut06}
\bibinfo{author}{\bibfnamefont{G.}~\bibnamefont{Talut}},
  \bibinfo{author}{\bibfnamefont{H.}~\bibnamefont{Reuther}},
  \bibinfo{author}{\bibfnamefont{A.}~\bibnamefont{M\"{u}cklich}},
  \bibinfo{author}{\bibfnamefont{F.}~\bibnamefont{Eichhorn}}, \bibnamefont{and}
  \bibinfo{author}{\bibfnamefont{K.}~\bibnamefont{Potzger}},
  \bibinfo{journal}{Appl. Phys. Lett.} \textbf{\bibinfo{volume}{89}},
  \bibinfo{pages}{161909} (\bibinfo{year}{2006}).

\bibitem[{\citenamefont{Zhou et~al.}(2006)\citenamefont{Zhou, Potzger, Zhang,
  Eichhorn, Skorupa, Helm, and Fassbender}}]{zhou06}
\bibinfo{author}{\bibfnamefont{S.}~\bibnamefont{Zhou}},
  \bibinfo{author}{\bibfnamefont{K.}~\bibnamefont{Potzger}},
  \bibinfo{author}{\bibfnamefont{G.}~\bibnamefont{Zhang}},
  \bibinfo{author}{\bibfnamefont{F.}~\bibnamefont{Eichhorn}},
  \bibinfo{author}{\bibfnamefont{W.}~\bibnamefont{Skorupa}},
  \bibinfo{author}{\bibfnamefont{M.}~\bibnamefont{Helm}}, \bibnamefont{and}
  \bibinfo{author}{\bibfnamefont{J.}~\bibnamefont{Fassbender}},
  \bibinfo{journal}{J. Appl. Phys.} \textbf{\bibinfo{volume}{100}},
  \bibinfo{pages}{114304} (\bibinfo{year}{2006}).

\bibitem[{\citenamefont{Zhou et~al.}(2007{\natexlab{a}})\citenamefont{Zhou,
  Potzger, Zhang, M\"{u}cklich, Eichhorn, Schell, Grotzschel, Schmidt, Skorupa,
  Helm et~al.}}]{zhou07JPD}
\bibinfo{author}{\bibfnamefont{S.}~\bibnamefont{Zhou}},
  \bibinfo{author}{\bibfnamefont{K.}~\bibnamefont{Potzger}},
  \bibinfo{author}{\bibfnamefont{G.}~\bibnamefont{Zhang}},
  \bibinfo{author}{\bibfnamefont{A.}~\bibnamefont{M\"{u}cklich}},
  \bibinfo{author}{\bibfnamefont{F.}~\bibnamefont{Eichhorn}},
  \bibinfo{author}{\bibfnamefont{N.}~\bibnamefont{Schell}},
  \bibinfo{author}{\bibfnamefont{R.}~\bibnamefont{Grotzschel}},
  \bibinfo{author}{\bibfnamefont{B.}~\bibnamefont{Schmidt}},
  \bibinfo{author}{\bibfnamefont{W.}~\bibnamefont{Skorupa}},
  \bibinfo{author}{\bibfnamefont{M.}~\bibnamefont{Helm}}, \bibnamefont{et~al.},
  \bibinfo{journal}{J. Phys. D-Appl. Phys.} \textbf{\bibinfo{volume}{40}},
  \bibinfo{pages}{964} (\bibinfo{year}{2007}{\natexlab{a}}).

\bibitem[{\citenamefont{Hebard et~al.}(2004)\citenamefont{Hebard, Rairigh,
  Kelly, Pearton, Abernathy, Chu, and Wilson}}]{hebard04}
\bibinfo{author}{\bibfnamefont{A.~F.} \bibnamefont{Hebard}},
  \bibinfo{author}{\bibfnamefont{R.~P.} \bibnamefont{Rairigh}},
  \bibinfo{author}{\bibfnamefont{J.~G.} \bibnamefont{Kelly}},
  \bibinfo{author}{\bibfnamefont{S.~J.} \bibnamefont{Pearton}},
  \bibinfo{author}{\bibfnamefont{C.~R.} \bibnamefont{Abernathy}},
  \bibinfo{author}{\bibfnamefont{S.~N.~G.} \bibnamefont{Chu}},
  \bibnamefont{and} \bibinfo{author}{\bibfnamefont{R.~G.}
  \bibnamefont{Wilson}}, \bibinfo{journal}{J. Phys. D-Appl. Phys.}
  \textbf{\bibinfo{volume}{37}}, \bibinfo{pages}{511} (\bibinfo{year}{2004}).

\bibitem[{\citenamefont{Ohshima et~al.}(2004)\citenamefont{Ohshima, Ogino,
  Niikura, Maeda, Sato, Ito, and Fukuda}}]{ohshima04}
\bibinfo{author}{\bibfnamefont{E.}~\bibnamefont{Ohshima}},
  \bibinfo{author}{\bibfnamefont{H.}~\bibnamefont{Ogino}},
  \bibinfo{author}{\bibfnamefont{I.}~\bibnamefont{Niikura}},
  \bibinfo{author}{\bibfnamefont{K.}~\bibnamefont{Maeda}},
  \bibinfo{author}{\bibfnamefont{M.}~\bibnamefont{Sato}},
  \bibinfo{author}{\bibfnamefont{M.}~\bibnamefont{Ito}}, \bibnamefont{and}
  \bibinfo{author}{\bibfnamefont{T.}~\bibnamefont{Fukuda}},
  \bibinfo{journal}{J. Cryst. Growth} \textbf{\bibinfo{volume}{260}},
  \bibinfo{pages}{166} (\bibinfo{year}{2004}).

\bibitem[{\citenamefont{Izyumskaya et~al.}(2007)\citenamefont{Izyumskaya,
  Avrutin, \"{O}zg\"{u}r, Alivov, and Morkoç}}]{izyumskaya07}
\bibinfo{author}{\bibfnamefont{N.}~\bibnamefont{Izyumskaya}},
  \bibinfo{author}{\bibfnamefont{V.}~\bibnamefont{Avrutin}},
  \bibinfo{author}{\bibfnamefont{U.}~\bibnamefont{\"{O}zg\"{u}r}},
  \bibinfo{author}{\bibfnamefont{Y.~I.} \bibnamefont{Alivov}},
  \bibnamefont{and} \bibinfo{author}{\bibfnamefont{H.}~\bibnamefont{Morkoç}},
  \bibinfo{journal}{Phys. Status Solidi B} \textbf{\bibinfo{volume}{244}},
  \bibinfo{pages}{1439} (\bibinfo{year}{2007}).

\bibitem[{\citenamefont{Endo et~al.}(2007)\citenamefont{Endo, Sugibuchi,
  Takahashi, Goto, Sugimura, Hane, and Kashiwaba}}]{endo:121906}
\bibinfo{author}{\bibfnamefont{H.}~\bibnamefont{Endo}},
  \bibinfo{author}{\bibfnamefont{M.}~\bibnamefont{Sugibuchi}},
  \bibinfo{author}{\bibfnamefont{K.}~\bibnamefont{Takahashi}},
  \bibinfo{author}{\bibfnamefont{S.}~\bibnamefont{Goto}},
  \bibinfo{author}{\bibfnamefont{S.}~\bibnamefont{Sugimura}},
  \bibinfo{author}{\bibfnamefont{K.}~\bibnamefont{Hane}}, \bibnamefont{and}
  \bibinfo{author}{\bibfnamefont{Y.}~\bibnamefont{Kashiwaba}},
  \bibinfo{journal}{Appl. Phys. Lett.} \textbf{\bibinfo{volume}{90}},
  \bibinfo{pages}{121906} (\bibinfo{year}{2007}).

\bibitem[{\citenamefont{Gao et~al.}(2006)\citenamefont{Gao, Zhang, Deng, Wang,
  Sun, and Zheng}}]{gao:123125}
\bibinfo{author}{\bibfnamefont{S.}~\bibnamefont{Gao}},
  \bibinfo{author}{\bibfnamefont{H.}~\bibnamefont{Zhang}},
  \bibinfo{author}{\bibfnamefont{R.}~\bibnamefont{Deng}},
  \bibinfo{author}{\bibfnamefont{X.}~\bibnamefont{Wang}},
  \bibinfo{author}{\bibfnamefont{D.}~\bibnamefont{Sun}}, \bibnamefont{and}
  \bibinfo{author}{\bibfnamefont{G.}~\bibnamefont{Zheng}},
  \bibinfo{journal}{Appl. Phys. Lett.} \textbf{\bibinfo{volume}{89}},
  \bibinfo{pages}{123125} (\bibinfo{year}{2006}).

\bibitem[{\citenamefont{Li et~al.}(2006)\citenamefont{Li, Kang, Lin, Chu, Chen,
  Wu, Yan, Chen, and Huang}}]{li:112507}
\bibinfo{author}{\bibfnamefont{W.}~\bibnamefont{Li}},
  \bibinfo{author}{\bibfnamefont{Q.}~\bibnamefont{Kang}},
  \bibinfo{author}{\bibfnamefont{Z.}~\bibnamefont{Lin}},
  \bibinfo{author}{\bibfnamefont{W.}~\bibnamefont{Chu}},
  \bibinfo{author}{\bibfnamefont{D.}~\bibnamefont{Chen}},
  \bibinfo{author}{\bibfnamefont{Z.}~\bibnamefont{Wu}},
  \bibinfo{author}{\bibfnamefont{Y.}~\bibnamefont{Yan}},
  \bibinfo{author}{\bibfnamefont{D.}~\bibnamefont{Chen}}, \bibnamefont{and}
  \bibinfo{author}{\bibfnamefont{F.}~\bibnamefont{Huang}},
  \bibinfo{journal}{Appl. Phys. Lett.} \textbf{\bibinfo{volume}{89}},
  \bibinfo{pages}{112507} (\bibinfo{year}{2006}).

\bibitem[{\citenamefont{Fenwick et~al.}(2007)\citenamefont{Fenwick, Kane,
  Varatharajan, Zaidi, Fang, Nemeth, Keeble, El-Mkami, Smith, Nause
  et~al.}}]{fenwick:64741Q}
\bibinfo{author}{\bibfnamefont{W.~E.} \bibnamefont{Fenwick}},
  \bibinfo{author}{\bibfnamefont{M.~H.} \bibnamefont{Kane}},
  \bibinfo{author}{\bibfnamefont{R.}~\bibnamefont{Varatharajan}},
  \bibinfo{author}{\bibfnamefont{T.}~\bibnamefont{Zaidi}},
  \bibinfo{author}{\bibfnamefont{Z.}~\bibnamefont{Fang}},
  \bibinfo{author}{\bibfnamefont{B.}~\bibnamefont{Nemeth}},
  \bibinfo{author}{\bibfnamefont{D.~J.} \bibnamefont{Keeble}},
  \bibinfo{author}{\bibfnamefont{H.}~\bibnamefont{El-Mkami}},
  \bibinfo{author}{\bibfnamefont{G.~M.} \bibnamefont{Smith}},
  \bibinfo{author}{\bibfnamefont{J.}~\bibnamefont{Nause}},
  \bibnamefont{et~al.}, \bibinfo{journal}{Proc. SPIE}
  \textbf{\bibinfo{volume}{6474}}, \bibinfo{pages}{64741Q}
  (\bibinfo{year}{2007}).

\bibitem[{\citenamefont{Matsui et~al.}(2004)\citenamefont{Matsui, Saeki, Kawai,
  Sasaki, Yoshimoto, Tsubaki, and Tabata}}]{matsui:2454}
\bibinfo{author}{\bibfnamefont{H.}~\bibnamefont{Matsui}},
  \bibinfo{author}{\bibfnamefont{H.}~\bibnamefont{Saeki}},
  \bibinfo{author}{\bibfnamefont{T.}~\bibnamefont{Kawai}},
  \bibinfo{author}{\bibfnamefont{A.}~\bibnamefont{Sasaki}},
  \bibinfo{author}{\bibfnamefont{M.}~\bibnamefont{Yoshimoto}},
  \bibinfo{author}{\bibfnamefont{M.}~\bibnamefont{Tsubaki}}, \bibnamefont{and}
  \bibinfo{author}{\bibfnamefont{H.}~\bibnamefont{Tabata}},
  \bibinfo{journal}{J. Vac. Sci. Technol. B} \textbf{\bibinfo{volume}{22}},
  \bibinfo{pages}{2454} (\bibinfo{year}{2004}).

\bibitem[{\citenamefont{Matsukura et~al.}(2002)\citenamefont{Matsukura, Ohno,
  and Dietl}}]{ohnohb}
\bibinfo{author}{\bibfnamefont{F.}~\bibnamefont{Matsukura}},
  \bibinfo{author}{\bibfnamefont{H.}~\bibnamefont{Ohno}}, \bibnamefont{and}
  \bibinfo{author}{\bibfnamefont{T.}~\bibnamefont{Dietl}},
  \emph{\bibinfo{title}{Handbook of Magnetic Materials}}
  (\bibinfo{publisher}{North-Holland}, \bibinfo{address}{Amsterdam},
  \bibinfo{year}{2002}).

\bibitem[{\citenamefont{Wellmann et~al.}(1998)\citenamefont{Wellmann, Garcia,
  Feng, and Petroff}}]{wellmann98}
\bibinfo{author}{\bibfnamefont{P.~J.} \bibnamefont{Wellmann}},
  \bibinfo{author}{\bibfnamefont{J.~M.} \bibnamefont{Garcia}},
  \bibinfo{author}{\bibfnamefont{J.~L.} \bibnamefont{Feng}}, \bibnamefont{and}
  \bibinfo{author}{\bibfnamefont{P.~M.} \bibnamefont{Petroff}},
  \bibinfo{journal}{Appl. Phys. Lett.} \textbf{\bibinfo{volume}{73}},
  \bibinfo{pages}{3291} (\bibinfo{year}{1998}).

\bibitem[{\citenamefont{Yuldashev et~al.}(2001)\citenamefont{Yuldashev, Shon,
  Kwon, Fu, Kim, Kim, Kang, and Fan}}]{yuldashev01}
\bibinfo{author}{\bibfnamefont{S.~U.} \bibnamefont{Yuldashev}},
  \bibinfo{author}{\bibfnamefont{Y.}~\bibnamefont{Shon}},
  \bibinfo{author}{\bibfnamefont{Y.~H.} \bibnamefont{Kwon}},
  \bibinfo{author}{\bibfnamefont{D.~J.} \bibnamefont{Fu}},
  \bibinfo{author}{\bibfnamefont{D.~Y.} \bibnamefont{Kim}},
  \bibinfo{author}{\bibfnamefont{H.~J.} \bibnamefont{Kim}},
  \bibinfo{author}{\bibfnamefont{T.~W.} \bibnamefont{Kang}}, \bibnamefont{and}
  \bibinfo{author}{\bibfnamefont{X.}~\bibnamefont{Fan}}, \bibinfo{journal}{J.
  Appl. Phys.} \textbf{\bibinfo{volume}{90}}, \bibinfo{pages}{3004}
  (\bibinfo{year}{2001}).

\bibitem[{\citenamefont{Ramsteiner et~al.}(2002)\citenamefont{Ramsteiner, Hao,
  Kawaharazuka, Zhu, Kastner, Hey, Daweritz, Grahn, and Ploog}}]{ramsteiner02}
\bibinfo{author}{\bibfnamefont{M.}~\bibnamefont{Ramsteiner}},
  \bibinfo{author}{\bibfnamefont{H.~Y.} \bibnamefont{Hao}},
  \bibinfo{author}{\bibfnamefont{A.}~\bibnamefont{Kawaharazuka}},
  \bibinfo{author}{\bibfnamefont{H.~J.} \bibnamefont{Zhu}},
  \bibinfo{author}{\bibfnamefont{M.}~\bibnamefont{Kastner}},
  \bibinfo{author}{\bibfnamefont{R.}~\bibnamefont{Hey}},
  \bibinfo{author}{\bibfnamefont{L.}~\bibnamefont{Daweritz}},
  \bibinfo{author}{\bibfnamefont{H.~T.} \bibnamefont{Grahn}}, \bibnamefont{and}
  \bibinfo{author}{\bibfnamefont{K.~H.} \bibnamefont{Ploog}},
  \bibinfo{journal}{Phys. Rev. B} \textbf{\bibinfo{volume}{66}},
  \bibinfo{pages}{4} (\bibinfo{year}{2002}).

\bibitem[{\citenamefont{Yokoyama et~al.}(2006)\citenamefont{Yokoyama, Ogawa,
  Nazmul, and Tanaka}}]{yokoyama06}
\bibinfo{author}{\bibfnamefont{M.}~\bibnamefont{Yokoyama}},
  \bibinfo{author}{\bibfnamefont{T.}~\bibnamefont{Ogawa}},
  \bibinfo{author}{\bibfnamefont{A.~M.} \bibnamefont{Nazmul}},
  \bibnamefont{and} \bibinfo{author}{\bibfnamefont{M.}~\bibnamefont{Tanaka}},
  \bibinfo{journal}{J. Appl. Phys.} \textbf{\bibinfo{volume}{99}},
  \bibinfo{pages}{08D502} (\bibinfo{year}{2006}).

\bibitem[{\citenamefont{Sato et~al.}(2005)\citenamefont{Sato, Katayama-Yoshida,
  and Dederichs}}]{sato05}
\bibinfo{author}{\bibfnamefont{K.}~\bibnamefont{Sato}},
  \bibinfo{author}{\bibfnamefont{H.}~\bibnamefont{Katayama-Yoshida}},
  \bibnamefont{and} \bibinfo{author}{\bibfnamefont{P.~H.}
  \bibnamefont{Dederichs}}, \bibinfo{journal}{Jpn. J. Appl. Phys.}
  \textbf{\bibinfo{volume}{44}}, \bibinfo{pages}{L948} (\bibinfo{year}{2005}).

\bibitem[{\citenamefont{Katayama-Yoshida
  et~al.}(2007)\citenamefont{Katayama-Yoshida, Sato, Fukushima, Toyoda, Kizaki,
  Dinh, and Dederichs}}]{katayama07}
\bibinfo{author}{\bibfnamefont{H.}~\bibnamefont{Katayama-Yoshida}},
  \bibinfo{author}{\bibfnamefont{K.}~\bibnamefont{Sato}},
  \bibinfo{author}{\bibfnamefont{T.}~\bibnamefont{Fukushima}},
  \bibinfo{author}{\bibfnamefont{M.}~\bibnamefont{Toyoda}},
  \bibinfo{author}{\bibfnamefont{H.}~\bibnamefont{Kizaki}},
  \bibinfo{author}{\bibfnamefont{V.~A.} \bibnamefont{Dinh}}, \bibnamefont{and}
  \bibinfo{author}{\bibfnamefont{P.~H.} \bibnamefont{Dederichs}},
  \bibinfo{journal}{Phys. Status Solidi. A} \textbf{\bibinfo{volume}{204}},
  \bibinfo{pages}{15} (\bibinfo{year}{2007}).

\bibitem[{\citenamefont{Liu et~al.}(2007)\citenamefont{Liu, Zhang, Shen, Wu,
  Li, Li, Lu, and Fan}}]{liu:092507}
\bibinfo{author}{\bibfnamefont{K.~W.} \bibnamefont{Liu}},
  \bibinfo{author}{\bibfnamefont{J.~Y.} \bibnamefont{Zhang}},
  \bibinfo{author}{\bibfnamefont{D.~Z.} \bibnamefont{Shen}},
  \bibinfo{author}{\bibfnamefont{X.~J.} \bibnamefont{Wu}},
  \bibinfo{author}{\bibfnamefont{B.~H.} \bibnamefont{Li}},
  \bibinfo{author}{\bibfnamefont{B.~S.} \bibnamefont{Li}},
  \bibinfo{author}{\bibfnamefont{Y.~M.} \bibnamefont{Lu}}, \bibnamefont{and}
  \bibinfo{author}{\bibfnamefont{X.~W.} \bibnamefont{Fan}},
  \bibinfo{journal}{Appl. Phys. Lett.} \textbf{\bibinfo{volume}{90}},
  \bibinfo{pages}{092507} (\bibinfo{year}{2007}).

\bibitem[{\citenamefont{Jamet et~al.}(2006)\citenamefont{Jamet, Barski,
  Devillers, Poydenot, Dujardin, Bayle-Guillemaud, Rothman, Bellet-Amalric,
  Marty, Cibert et~al.}}]{jamet06}
\bibinfo{author}{\bibfnamefont{M.}~\bibnamefont{Jamet}},
  \bibinfo{author}{\bibfnamefont{A.}~\bibnamefont{Barski}},
  \bibinfo{author}{\bibfnamefont{T.}~\bibnamefont{Devillers}},
  \bibinfo{author}{\bibfnamefont{V.}~\bibnamefont{Poydenot}},
  \bibinfo{author}{\bibfnamefont{R.}~\bibnamefont{Dujardin}},
  \bibinfo{author}{\bibfnamefont{P.}~\bibnamefont{Bayle-Guillemaud}},
  \bibinfo{author}{\bibfnamefont{J.}~\bibnamefont{Rothman}},
  \bibinfo{author}{\bibfnamefont{E.}~\bibnamefont{Bellet-Amalric}},
  \bibinfo{author}{\bibfnamefont{A.}~\bibnamefont{Marty}},
  \bibinfo{author}{\bibfnamefont{J.}~\bibnamefont{Cibert}},
  \bibnamefont{et~al.}, \bibinfo{journal}{Nat. Mater.}
  \textbf{\bibinfo{volume}{5}}, \bibinfo{pages}{653} (\bibinfo{year}{2006}).

\bibitem[{\citenamefont{Dietl and Ohno}(2006)}]{dietlmat}
\bibinfo{author}{\bibfnamefont{T.}~\bibnamefont{Dietl}} \bibnamefont{and}
  \bibinfo{author}{\bibfnamefont{H.}~\bibnamefont{Ohno}},
  \bibinfo{journal}{Materials Today} \textbf{\bibinfo{volume}{9}},
  \bibinfo{pages}{18} (\bibinfo{year}{2006}).

\bibitem[{\citenamefont{Ziegler et~al.}(1985)\citenamefont{Ziegler, Biersack,
  and Littmark}}]{trim}
\bibinfo{author}{\bibfnamefont{J.}~\bibnamefont{Ziegler}},
  \bibinfo{author}{\bibfnamefont{J.}~\bibnamefont{Biersack}}, \bibnamefont{and}
  \bibinfo{author}{\bibfnamefont{U.}~\bibnamefont{Littmark}},
  \emph{\bibinfo{title}{The stopping and range of ions in matter}}
  (\bibinfo{publisher}{Pergamon}, \bibinfo{address}{New York},
  \bibinfo{year}{1985}).

\bibitem[{\citenamefont{Chu et~al.}(1978)\citenamefont{Chu, Mayer, and
  Nicolet}}]{chuwk}
\bibinfo{author}{\bibfnamefont{W.~K.} \bibnamefont{Chu}},
  \bibinfo{author}{\bibfnamefont{J.~W.} \bibnamefont{Mayer}}, \bibnamefont{and}
  \bibinfo{author}{\bibfnamefont{M.~A.} \bibnamefont{Nicolet}},
  \emph{\bibinfo{title}{Backscattering Spectrometry}}
  (\bibinfo{publisher}{Academic}, \bibinfo{address}{New York},
  \bibinfo{year}{1978}).

\bibitem[{\citenamefont{Brand}(1987)}]{brand87}
\bibinfo{author}{\bibfnamefont{R.}~\bibnamefont{Brand}},
  \bibinfo{journal}{Nucl. Instrum. Methods Phys. Res.}
  \textbf{\bibinfo{volume}{B 28}}, \bibinfo{pages}{398} (\bibinfo{year}{1987}).

\bibitem[{\citenamefont{Kucheyev et~al.}(2003)\citenamefont{Kucheyev, Williams,
  Jagadish, Zou, Evans, Nelson, and Hamza}}]{kucheyev03}
\bibinfo{author}{\bibfnamefont{S.~O.} \bibnamefont{Kucheyev}},
  \bibinfo{author}{\bibfnamefont{J.~S.} \bibnamefont{Williams}},
  \bibinfo{author}{\bibfnamefont{C.}~\bibnamefont{Jagadish}},
  \bibinfo{author}{\bibfnamefont{J.}~\bibnamefont{Zou}},
  \bibinfo{author}{\bibfnamefont{C.}~\bibnamefont{Evans}},
  \bibinfo{author}{\bibfnamefont{A.~J.} \bibnamefont{Nelson}},
  \bibnamefont{and} \bibinfo{author}{\bibfnamefont{A.~V.} \bibnamefont{Hamza}},
  \bibinfo{journal}{Phys. Rev. B} \textbf{\bibinfo{volume}{67}},
  \bibinfo{pages}{094115} (\bibinfo{year}{2003}).

\bibitem[{\citenamefont{Kucheyev et~al.}(2002)\citenamefont{Kucheyev,
  Deenapanray, Jagadish, Williams, Yano, Koike, Sasa, Inoue, and ichi
  Ogata}}]{kucheyev02}
\bibinfo{author}{\bibfnamefont{S.~O.} \bibnamefont{Kucheyev}},
  \bibinfo{author}{\bibfnamefont{P.~N.~K.} \bibnamefont{Deenapanray}},
  \bibinfo{author}{\bibfnamefont{C.}~\bibnamefont{Jagadish}},
  \bibinfo{author}{\bibfnamefont{J.~S.} \bibnamefont{Williams}},
  \bibinfo{author}{\bibfnamefont{M.}~\bibnamefont{Yano}},
  \bibinfo{author}{\bibfnamefont{K.}~\bibnamefont{Koike}},
  \bibinfo{author}{\bibfnamefont{S.}~\bibnamefont{Sasa}},
  \bibinfo{author}{\bibfnamefont{M.}~\bibnamefont{Inoue}}, \bibnamefont{and}
  \bibinfo{author}{\bibfnamefont{K.}~\bibnamefont{ichi Ogata}},
  \bibinfo{journal}{Appl. Phys. Lett.} \textbf{\bibinfo{volume}{81}},
  \bibinfo{pages}{3350} (\bibinfo{year}{2002}).

\bibitem[{\citenamefont{Kucheyev et~al.}(2001)\citenamefont{Kucheyev, Williams,
  and Pearton}}]{kucheyev01}
\bibinfo{author}{\bibfnamefont{S.~O.} \bibnamefont{Kucheyev}},
  \bibinfo{author}{\bibfnamefont{J.~S.} \bibnamefont{Williams}},
  \bibnamefont{and} \bibinfo{author}{\bibfnamefont{S.~J.}
  \bibnamefont{Pearton}}, \bibinfo{journal}{Mater. Sci. Eng.}
  \textbf{\bibinfo{volume}{R33}}, \bibinfo{pages}{51} (\bibinfo{year}{2001}).

\bibitem[{\citenamefont{Coleman et~al.}(2005)\citenamefont{Coleman, Tan,
  Jagadish, Kucheyev, and Zou}}]{coleman:231912}
\bibinfo{author}{\bibfnamefont{V.~A.} \bibnamefont{Coleman}},
  \bibinfo{author}{\bibfnamefont{H.~H.} \bibnamefont{Tan}},
  \bibinfo{author}{\bibfnamefont{C.}~\bibnamefont{Jagadish}},
  \bibinfo{author}{\bibfnamefont{S.~O.} \bibnamefont{Kucheyev}},
  \bibnamefont{and} \bibinfo{author}{\bibfnamefont{J.}~\bibnamefont{Zou}},
  \bibinfo{journal}{Appl. Phys. Lett.} \textbf{\bibinfo{volume}{87}},
  \bibinfo{pages}{231912} (\bibinfo{year}{2005}).

\bibitem[{\citenamefont{Cullity}(1978)}]{scherrer}
\bibinfo{author}{\bibfnamefont{B.~D.} \bibnamefont{Cullity}},
  \emph{\bibinfo{title}{Elements of X-ray Diffractions}}
  (\bibinfo{publisher}{Reading}, \bibinfo{address}{Addison-Wesley},
  \bibinfo{year}{1978}).

\bibitem[{\citenamefont{Zhou et~al.}(2007{\natexlab{b}})\citenamefont{Zhou,
  Potzger, von Borany, Gr\"{o}tschel, Skorupa, Helm, and
  Fassbender}}]{zhou07CoNi}
\bibinfo{author}{\bibfnamefont{S.}~\bibnamefont{Zhou}},
  \bibinfo{author}{\bibfnamefont{K.}~\bibnamefont{Potzger}},
  \bibinfo{author}{\bibfnamefont{J.}~\bibnamefont{von Borany}},
  \bibinfo{author}{\bibfnamefont{R.}~\bibnamefont{Gr\"{o}tschel}},
  \bibinfo{author}{\bibfnamefont{W.}~\bibnamefont{Skorupa}},
  \bibinfo{author}{\bibfnamefont{M.}~\bibnamefont{Helm}}, \bibnamefont{and}
  \bibinfo{author}{\bibfnamefont{J.}~\bibnamefont{Fassbender}}
  (\bibinfo{year}{2007}{\natexlab{b}}), \bibinfo{note}{in preparation}.

\bibitem[{\citenamefont{Liu et~al.}(1997)\citenamefont{Liu, Mensching, Volz,
  and Rauschenbach}}]{GaN_expension}
\bibinfo{author}{\bibfnamefont{C.}~\bibnamefont{Liu}},
  \bibinfo{author}{\bibfnamefont{B.}~\bibnamefont{Mensching}},
  \bibinfo{author}{\bibfnamefont{K.}~\bibnamefont{Volz}}, \bibnamefont{and}
  \bibinfo{author}{\bibfnamefont{B.}~\bibnamefont{Rauschenbach}},
  \bibinfo{journal}{Appl. Phys. Lett.} \textbf{\bibinfo{volume}{71}},
  \bibinfo{pages}{2313} (\bibinfo{year}{1997}).

\bibitem[{\citenamefont{Ronning et~al.}(2001)\citenamefont{Ronning, Carlsonb,
  and Davis}}]{GaN_expension_review}
\bibinfo{author}{\bibfnamefont{C.}~\bibnamefont{Ronning}},
  \bibinfo{author}{\bibfnamefont{E.~P.} \bibnamefont{Carlsonb}},
  \bibnamefont{and} \bibinfo{author}{\bibfnamefont{R.~F.} \bibnamefont{Davis}},
  \bibinfo{journal}{Phys. Rep.} \textbf{\bibinfo{volume}{351}},
  \bibinfo{pages}{349} (\bibinfo{year}{2001}).

\bibitem[{\citenamefont{Respaud et~al.}(1998)\citenamefont{Respaud, Broto,
  Rakoto, Fert, Thomas, Barbara, Verelst, Snoeck, Lecante, Mosset
  et~al.}}]{respaud}
\bibinfo{author}{\bibfnamefont{M.}~\bibnamefont{Respaud}},
  \bibinfo{author}{\bibfnamefont{J.~M.} \bibnamefont{Broto}},
  \bibinfo{author}{\bibfnamefont{H.}~\bibnamefont{Rakoto}},
  \bibinfo{author}{\bibfnamefont{A.~R.} \bibnamefont{Fert}},
  \bibinfo{author}{\bibfnamefont{L.}~\bibnamefont{Thomas}},
  \bibinfo{author}{\bibfnamefont{B.}~\bibnamefont{Barbara}},
  \bibinfo{author}{\bibfnamefont{M.}~\bibnamefont{Verelst}},
  \bibinfo{author}{\bibfnamefont{E.}~\bibnamefont{Snoeck}},
  \bibinfo{author}{\bibfnamefont{P.}~\bibnamefont{Lecante}},
  \bibinfo{author}{\bibfnamefont{A.}~\bibnamefont{Mosset}},
  \bibnamefont{et~al.}, \bibinfo{journal}{Phys. Rev. B}
  \textbf{\bibinfo{volume}{57}}, \bibinfo{pages}{2925} (\bibinfo{year}{1998}).

\bibitem[{\citenamefont{Shinde et~al.}(2004)\citenamefont{Shinde, Ogale,
  Higgins, Zheng, Millis, Kulkarni, Ramesh, Greene, and Venkatesan}}]{shinde04}
\bibinfo{author}{\bibfnamefont{S.~R.} \bibnamefont{Shinde}},
  \bibinfo{author}{\bibfnamefont{S.~B.} \bibnamefont{Ogale}},
  \bibinfo{author}{\bibfnamefont{J.~S.} \bibnamefont{Higgins}},
  \bibinfo{author}{\bibfnamefont{H.}~\bibnamefont{Zheng}},
  \bibinfo{author}{\bibfnamefont{A.~J.} \bibnamefont{Millis}},
  \bibinfo{author}{\bibfnamefont{V.~N.} \bibnamefont{Kulkarni}},
  \bibinfo{author}{\bibfnamefont{R.}~\bibnamefont{Ramesh}},
  \bibinfo{author}{\bibfnamefont{R.~L.} \bibnamefont{Greene}},
  \bibnamefont{and}
  \bibinfo{author}{\bibfnamefont{T.}~\bibnamefont{Venkatesan}},
  \bibinfo{journal}{Phys. Rev. Lett.} \textbf{\bibinfo{volume}{92}},
  \bibinfo{pages}{166601} (\bibinfo{year}{2004}).

\bibitem[{\citenamefont{Farle}(2005)}]{farle}
\bibinfo{author}{\bibfnamefont{M.}~\bibnamefont{Farle}}, in
  \emph{\bibinfo{booktitle}{Magnetism goes Nano}} (\bibinfo{address}{Juelich,
  Germany}, \bibinfo{year}{2005}), p. \bibinfo{pages}{C4.2}.

\bibitem[{\citenamefont{Sun et~al.}(2003)\citenamefont{Sun, Salamon, Garnier,
  and Averback}}]{PhysRevLett.91.167206}
\bibinfo{author}{\bibfnamefont{Y.}~\bibnamefont{Sun}},
  \bibinfo{author}{\bibfnamefont{M.~B.} \bibnamefont{Salamon}},
  \bibinfo{author}{\bibfnamefont{K.}~\bibnamefont{Garnier}}, \bibnamefont{and}
  \bibinfo{author}{\bibfnamefont{R.~S.} \bibnamefont{Averback}},
  \bibinfo{journal}{Phys. Rev. Lett.} \textbf{\bibinfo{volume}{91}},
  \bibinfo{pages}{167206} (\bibinfo{year}{2003}).

\bibitem[{\citenamefont{Tsoi et~al.}(2005)\citenamefont{Tsoi, Wenger,
  Senaratne, Tackett, Buc, Naik, Vaishnava, and Naik}}]{tsoi:014445}
\bibinfo{author}{\bibfnamefont{G.~M.} \bibnamefont{Tsoi}},
  \bibinfo{author}{\bibfnamefont{L.~E.} \bibnamefont{Wenger}},
  \bibinfo{author}{\bibfnamefont{U.}~\bibnamefont{Senaratne}},
  \bibinfo{author}{\bibfnamefont{R.~J.} \bibnamefont{Tackett}},
  \bibinfo{author}{\bibfnamefont{E.~C.} \bibnamefont{Buc}},
  \bibinfo{author}{\bibfnamefont{R.}~\bibnamefont{Naik}},
  \bibinfo{author}{\bibfnamefont{P.~P.} \bibnamefont{Vaishnava}},
  \bibnamefont{and} \bibinfo{author}{\bibfnamefont{V.}~\bibnamefont{Naik}},
  \bibinfo{journal}{Phys. Rev. B} \textbf{\bibinfo{volume}{72}},
  \bibinfo{pages}{014445} (\bibinfo{year}{2005}).

\bibitem[{\citenamefont{Jacobsohn et~al.}(2006)\citenamefont{Jacobsohn,
  Hundley, Thompson, Dickerson, and Nastasi}}]{jacobsohn:321}
\bibinfo{author}{\bibfnamefont{L.~G.} \bibnamefont{Jacobsohn}},
  \bibinfo{author}{\bibfnamefont{M.~F.} \bibnamefont{Hundley}},
  \bibinfo{author}{\bibfnamefont{J.~D.} \bibnamefont{Thompson}},
  \bibinfo{author}{\bibfnamefont{R.~M.} \bibnamefont{Dickerson}},
  \bibnamefont{and} \bibinfo{author}{\bibfnamefont{M.}~\bibnamefont{Nastasi}},
  \bibinfo{journal}{J. Vac. Sci. Technol. B} \textbf{\bibinfo{volume}{24}},
  \bibinfo{pages}{321} (\bibinfo{year}{2006}).

\bibitem[{\citenamefont{Batlle and Labarta}(2002)}]{Battle}
\bibinfo{author}{\bibfnamefont{X.}~\bibnamefont{Batlle}} \bibnamefont{and}
  \bibinfo{author}{\bibfnamefont{A.}~\bibnamefont{Labarta}},
  \bibinfo{journal}{J. Phys. D-Appl. Phys.} \textbf{\bibinfo{volume}{35}},
  \bibinfo{pages}{R15} (\bibinfo{year}{2002}).

\bibitem[{\citenamefont{Shaw et~al.}(2006)\citenamefont{Shaw, Lee, and
  Falco}}]{shaw:094417}
\bibinfo{author}{\bibfnamefont{J.~M.} \bibnamefont{Shaw}},
  \bibinfo{author}{\bibfnamefont{S.}~\bibnamefont{Lee}}, \bibnamefont{and}
  \bibinfo{author}{\bibfnamefont{C.~M.} \bibnamefont{Falco}},
  \bibinfo{journal}{Phys. Rev. B} \textbf{\bibinfo{volume}{73}},
  \bibinfo{pages}{094417} (\bibinfo{year}{2006}).

\bibitem[{\citenamefont{Pulwey et~al.}(2002)\citenamefont{Pulwey, Zolfl,
  Bayreuther, and Weiss}}]{pulwey:7995}
\bibinfo{author}{\bibfnamefont{R.}~\bibnamefont{Pulwey}},
  \bibinfo{author}{\bibfnamefont{M.}~\bibnamefont{Zolfl}},
  \bibinfo{author}{\bibfnamefont{G.}~\bibnamefont{Bayreuther}},
  \bibnamefont{and} \bibinfo{author}{\bibfnamefont{D.}~\bibnamefont{Weiss}},
  \bibinfo{journal}{J. Appl. Phys.} \textbf{\bibinfo{volume}{91}},
  \bibinfo{pages}{7995} (\bibinfo{year}{2002}).

\bibitem[{\citenamefont{Zheng et~al.}(2004)\citenamefont{Zheng, Gu, and
  Zhang}}]{zheng:139702}
\bibinfo{author}{\bibfnamefont{R.~K.} \bibnamefont{Zheng}},
  \bibinfo{author}{\bibfnamefont{H.}~\bibnamefont{Gu}}, \bibnamefont{and}
  \bibinfo{author}{\bibfnamefont{X.~X.} \bibnamefont{Zhang}},
  \bibinfo{journal}{Phys. Rev. Lett.} \textbf{\bibinfo{volume}{93}},
  \bibinfo{eid}{139702} (\bibinfo{year}{2004}).

\bibitem[{\citenamefont{Sasaki et~al.}(2005)\citenamefont{Sasaki, Jonsson,
  Takayama, and Mamiya}}]{sasaki:104405}
\bibinfo{author}{\bibfnamefont{M.}~\bibnamefont{Sasaki}},
  \bibinfo{author}{\bibfnamefont{P.~E.} \bibnamefont{Jonsson}},
  \bibinfo{author}{\bibfnamefont{H.}~\bibnamefont{Takayama}}, \bibnamefont{and}
  \bibinfo{author}{\bibfnamefont{H.}~\bibnamefont{Mamiya}},
  \bibinfo{journal}{Phys. Rev. B} \textbf{\bibinfo{volume}{71}},
  \bibinfo{pages}{104405} (\bibinfo{year}{2005}).

\bibitem[{\citenamefont{Chakraverty et~al.}(2006)\citenamefont{Chakraverty,
  Ghosh, Kumar, and Frydman}}]{chakraverty:042501}
\bibinfo{author}{\bibfnamefont{S.}~\bibnamefont{Chakraverty}},
  \bibinfo{author}{\bibfnamefont{B.}~\bibnamefont{Ghosh}},
  \bibinfo{author}{\bibfnamefont{S.}~\bibnamefont{Kumar}}, \bibnamefont{and}
  \bibinfo{author}{\bibfnamefont{A.}~\bibnamefont{Frydman}},
  \bibinfo{journal}{Appl. Phys. Lett.} \textbf{\bibinfo{volume}{88}},
  \bibinfo{pages}{042501} (\bibinfo{year}{2006}).

\bibitem[{\citenamefont{Kaidashev et~al.}(2003)\citenamefont{Kaidashev, Lorenz,
  von Wenckstern, Rahm, Semmelhack, Han, Benndorf, Bundesmann, Hochmuth, and
  Grundmann}}]{lorenz}
\bibinfo{author}{\bibfnamefont{E.~M.} \bibnamefont{Kaidashev}},
  \bibinfo{author}{\bibfnamefont{M.}~\bibnamefont{Lorenz}},
  \bibinfo{author}{\bibfnamefont{H.}~\bibnamefont{von Wenckstern}},
  \bibinfo{author}{\bibfnamefont{A.}~\bibnamefont{Rahm}},
  \bibinfo{author}{\bibfnamefont{H.-C.} \bibnamefont{Semmelhack}},
  \bibinfo{author}{\bibfnamefont{K.-H.} \bibnamefont{Han}},
  \bibinfo{author}{\bibfnamefont{G.}~\bibnamefont{Benndorf}},
  \bibinfo{author}{\bibfnamefont{C.}~\bibnamefont{Bundesmann}},
  \bibinfo{author}{\bibfnamefont{H.}~\bibnamefont{Hochmuth}}, \bibnamefont{and}
  \bibinfo{author}{\bibfnamefont{M.}~\bibnamefont{Grundmann}},
  \bibinfo{journal}{Appl. Phys. Lett.} \textbf{\bibinfo{volume}{82}},
  \bibinfo{pages}{3901} (\bibinfo{year}{2003}).

\bibitem[{\citenamefont{Lorenz et~al.}(2003)\citenamefont{Lorenz, Kaidashev,
  von Wenckstern, V., Bundesmann, Spemann, Benndorf, Hochmuth, Rahm, Semmelhack
  et~al.}}]{lorenz03}
\bibinfo{author}{\bibfnamefont{M.}~\bibnamefont{Lorenz}},
  \bibinfo{author}{\bibfnamefont{E.~M.} \bibnamefont{Kaidashev}},
  \bibinfo{author}{\bibfnamefont{H.}~\bibnamefont{von Wenckstern}},
  \bibinfo{author}{\bibfnamefont{R.}~\bibnamefont{V.}},
  \bibinfo{author}{\bibfnamefont{C.}~\bibnamefont{Bundesmann}},
  \bibinfo{author}{\bibfnamefont{D.}~\bibnamefont{Spemann}},
  \bibinfo{author}{\bibfnamefont{G.}~\bibnamefont{Benndorf}},
  \bibinfo{author}{\bibfnamefont{H.}~\bibnamefont{Hochmuth}},
  \bibinfo{author}{\bibfnamefont{A.}~\bibnamefont{Rahm}},
  \bibinfo{author}{\bibfnamefont{H.-C.} \bibnamefont{Semmelhack}},
  \bibnamefont{et~al.}, \bibinfo{journal}{Solid State Electron}
  \textbf{\bibinfo{volume}{47}}, \bibinfo{pages}{2205} (\bibinfo{year}{2003}).

\bibitem[{\citenamefont{Lorenz et~al.}(2006)\citenamefont{Lorenz, Johne, Nobis,
  Hochmuth, Lenzner, Grundmann, Schenk, Borenstain, Schon, Bekeny
  et~al.}}]{lorenz:243510}
\bibinfo{author}{\bibfnamefont{M.}~\bibnamefont{Lorenz}},
  \bibinfo{author}{\bibfnamefont{R.}~\bibnamefont{Johne}},
  \bibinfo{author}{\bibfnamefont{T.}~\bibnamefont{Nobis}},
  \bibinfo{author}{\bibfnamefont{H.}~\bibnamefont{Hochmuth}},
  \bibinfo{author}{\bibfnamefont{J.}~\bibnamefont{Lenzner}},
  \bibinfo{author}{\bibfnamefont{M.}~\bibnamefont{Grundmann}},
  \bibinfo{author}{\bibfnamefont{H.~P.~D.} \bibnamefont{Schenk}},
  \bibinfo{author}{\bibfnamefont{S.~I.} \bibnamefont{Borenstain}},
  \bibinfo{author}{\bibfnamefont{A.}~\bibnamefont{Schon}},
  \bibinfo{author}{\bibfnamefont{C.}~\bibnamefont{Bekeny}},
  \bibnamefont{et~al.}, \bibinfo{journal}{Appl. Phys. Lett.}
  \textbf{\bibinfo{volume}{89}}, \bibinfo{pages}{243510}
  (\bibinfo{year}{2006}).

\bibitem[{\citenamefont{Dietl}(2006)}]{dietlnatm}
\bibinfo{author}{\bibfnamefont{T.}~\bibnamefont{Dietl}}, \bibinfo{journal}{Nat.
  Mater.} \textbf{\bibinfo{volume}{5}}, \bibinfo{pages}{673}
  (\bibinfo{year}{2006}).

\bibitem[{\citenamefont{Narayan et~al.}(1998)\citenamefont{Narayan, Dovidenko,
  Sharma, and Oktyabrsky}}]{narayan:2597}
\bibinfo{author}{\bibfnamefont{J.}~\bibnamefont{Narayan}},
  \bibinfo{author}{\bibfnamefont{K.}~\bibnamefont{Dovidenko}},
  \bibinfo{author}{\bibfnamefont{A.~K.} \bibnamefont{Sharma}},
  \bibnamefont{and}
  \bibinfo{author}{\bibfnamefont{S.}~\bibnamefont{Oktyabrsky}},
  \bibinfo{journal}{J. Appl. Phys.} \textbf{\bibinfo{volume}{84}},
  \bibinfo{pages}{2597} (\bibinfo{year}{1998}).

\bibitem[{\citenamefont{Kaiser et~al.}(2002)\citenamefont{Kaiser, Muller,
  Grazul, Chuvilin, and Kawasaki}}]{Kaiser02}
\bibinfo{author}{\bibfnamefont{U.}~\bibnamefont{Kaiser}},
  \bibinfo{author}{\bibfnamefont{D.}~\bibnamefont{Muller}},
  \bibinfo{author}{\bibfnamefont{J.}~\bibnamefont{Grazul}},
  \bibinfo{author}{\bibfnamefont{A.}~\bibnamefont{Chuvilin}}, \bibnamefont{and}
  \bibinfo{author}{\bibfnamefont{M.}~\bibnamefont{Kawasaki}},
  \bibinfo{journal}{Nat. Mater.} \textbf{\bibinfo{volume}{1}},
  \bibinfo{pages}{102} (\bibinfo{year}{2002}).

\bibitem[{\citenamefont{Potzger et~al.}(2007)\citenamefont{Potzger, Anwand,
  Reuther, Zhou, Talut, Brauer, Skorupa, and Fassbender}}]{potzger07}
\bibinfo{author}{\bibfnamefont{K.}~\bibnamefont{Potzger}},
  \bibinfo{author}{\bibfnamefont{W.}~\bibnamefont{Anwand}},
  \bibinfo{author}{\bibfnamefont{H.}~\bibnamefont{Reuther}},
  \bibinfo{author}{\bibfnamefont{S.}~\bibnamefont{Zhou}},
  \bibinfo{author}{\bibfnamefont{G.}~\bibnamefont{Talut}},
  \bibinfo{author}{\bibfnamefont{G.}~\bibnamefont{Brauer}},
  \bibinfo{author}{\bibfnamefont{W.}~\bibnamefont{Skorupa}}, \bibnamefont{and}
  \bibinfo{author}{\bibfnamefont{J.}~\bibnamefont{Fassbender}},
  \bibinfo{journal}{J. Appl. Phys.} \textbf{\bibinfo{volume}{101}},
  \bibinfo{pages}{033906} (\bibinfo{year}{2007}).

\bibitem[{\citenamefont{Kolesnik et~al.}(2004)\citenamefont{Kolesnik,
  Dabrowski, and Mais}}]{kolesnik:2582}
\bibinfo{author}{\bibfnamefont{S.}~\bibnamefont{Kolesnik}},
  \bibinfo{author}{\bibfnamefont{B.}~\bibnamefont{Dabrowski}},
  \bibnamefont{and} \bibinfo{author}{\bibfnamefont{J.}~\bibnamefont{Mais}},
  \bibinfo{journal}{J. Appl. Phys.} \textbf{\bibinfo{volume}{95}},
  \bibinfo{pages}{2582} (\bibinfo{year}{2004}).

\end{thebibliography}
\end{document}